\documentclass[SIADSpaper,fleqn]{siamonline220329}

\usepackage{amsfonts}
\usepackage{graphicx}
\usepackage{epstopdf}

\usepackage{nicefrac}

\usepackage{algorithmic}
\ifpdf
  \DeclareGraphicsExtensions{.eps,.pdf,.png,.jpg}
\else
  \DeclareGraphicsExtensions{.eps}
\fi

\usepackage{enumitem}
\setlist[enumerate]{leftmargin=.5in}
\setlist[itemize]{leftmargin=.5in}

\newsiamremark{remark}{Remark}
\newsiamremark{hypothesis}{Hypothesis}
\crefname{hypothesis}{Hypothesis}{Hypotheses}
\newsiamthm{claim}{Claim}

\title{Heterogeneous populations of quadratic integrate-and-fire neurons:\\on the generality of Lorentzian distributions 
}

\author{Bastian Pietras, Ernest Montbri\'o\thanks{Department of Information and Communication Technologies,Universitat Pompeu Fabra, C/. \!T\`anger 122-140, 08018 Barcelona, Spain (\email{bastian.pietras@upf.edu}, \email{ernest.montbrio@upf.edu}).
}
}

\date{\today}

\usepackage{amsopn}

\ifpdf
\hypersetup{
  pdftitle={Heterogeneous populations of QIF neurons},
  pdfauthor={B. Pietras}
}
\fi

\begin{document}

\maketitle

\begin{abstract}
Over the last decade, next-generation neural mass models have become increasingly prominent in mathematical neuroscience.
These models link microscopic dynamics with low-dimensional systems of so-called firing rate equations that exactly capture the collective dynamics of large populations of heterogeneous quadratic integrate-and-fire (QIF) neurons.
A particularly tractable type of heterogeneity is the distribution of the QIF neurons' excitability parameters, or inputs, according to a Lorentzian.
While other distributions---such as those approximating Gaussian or uniform distributions---admit to exact mean-field reductions, they result in more complex firing rate equations that are challenging to analyze, and it remains unclear whether they produce comparable collective dynamics.
Here, we first demonstrate why Lorentzian heterogeneity is analytically favorable and, second, identify when it leads to qualitatively different collective dynamics compared to other types of heterogeneity.
A stationary mean-field approach enables us to derive explicit expressions for the distributions of the neurons' firing rates and voltages in macroscopic stationary states with arbitrary heterogeneities.
We also explicate the exclusive relationship between Lorentzian distributed inputs and Lorentzian distributed voltages, whose width happens to coincide with the population firing rate.
A dynamic mean-field approach for unimodal heterogeneities further allows us to comprehensively analyze and compare collective dynamics.
We find that different types of heterogeneity typically yield qualitatively similar dynamics. However, when gap junction coupling is present, Lorentzian heterogeneity induces nonuniversal behavior, obscuring a diversity-induced transition to synchrony.
\end{abstract}


\vspace{-1em}

\section{Introduction}
Population activity of spiking neurons in the brain is crucial to theoretical neuroscience as it reflects the complex and dynamic interactions within neural networks. Understanding how diverse neurons collectively contribute to information processing allows researchers to develop comprehensive models of cognitive functions and behaviors.
Given that no two neurons in our brain are identical, it seems appropriate to assume neuronal heterogeneity where individual dynamics differ from neuron to neuron.
But, how does heterogeneity on a neuron level affect the collective dynamics on a population level? Can we capture macroscopic effects of neuronal heterogeneity by means of mean-field analysis? And do different kinds of heterogeneity yield similar collective behavior, or does one type make a network more prone to, e.g., synchrony than others?
To answer these and related questions, researchers have employed bottom-up approches to bridge scales from microscopic, spiking neuron networks to macroscopic, collective activity \cite{tsodyks1993pattern,golomb1993dynamics,wang1996gamma,amit1997dynamics,chow1998phase,white1998synchronization,tiesinga2000robust,hansel2001existence,hansel2003asynchronous,jin2002fast,renart2003robust,assisi2005synchrony,shamir2006implications,gigante2007diverse,bathellier2008gamma,stefanescu2008low,
vladimirski2008episodic,ostojic2009synchronization,lafuerza2010nonuniversal,
luccioli2010irregular,roxin2011role,hermann2012heterogeneous,mejias2012optimal,mejias2014differential,nicola2013mean,yim2013impact,
di2014heterogeneous,ly2014dynamics,montbrio2015macroscopic,chandra2017modeling,nykamp2017mean,barreiro2018investigating,vegue2019firing,Daffertshofer2020,
adam2020inferring,baravalle2021heterogeneity,di2021optimal,laing2021effects,perez2021neural,
zeldenrust2021efficient,chen2022exact,guerreiro2022exact,gast2023macroscopic,gast2024neural,hutt2023intrinsic,hutt2024diversity}. 
Prototypical, and mathematically accessible, spiking neuron models are integrate-and-fire neurons, which have been used to study collective dynamics for heterogeneous networks of leaky \cite{tsodyks1993pattern,amit1997dynamics,ostojic2009synchronization,luccioli2010irregular,
mejias2012optimal,mejias2014differential,yim2013impact,vegue2019firing,adam2020inferring,di2021optimal} or quadratic integrate-and-fire neurons \cite{hansel2001existence,hansel2003asynchronous,bathellier2008gamma,nicola2013mean,montbrio2015macroscopic,laing2021effects,chen2022exact,guerreiro2022exact,gast2023macroscopic}.
As a disclaimer, here we consider neuronal heterogeneity in globally coupled networks by varying the neurons' excitability parameters, or equivalently, by distributing their input currents; we do not consider heterogeneity induced through, e.g., particular connectivity structures or distributed time constants.

Theoretical advances in the last decade have boosted the macroscopic description for large populations of quadratic integrate-and-fire (QIF) neurons, which inspired a novel class of so-called ``next-generation'' neural mass models~\cite{montbrio2015macroscopic,luke2013complete,laing2014derivation,coombes2018next,coombes2023next}.
Specifically, in~\cite{montbrio2015macroscopic}
Montbri\'o, Paz\'o and Roxin derived a two-dimensional system of ``firing rate equations'' that exactly capture the collective dynamics of globally coupled QIF neurons with Lorentzian heterogeneity.
A key ingredient for their success was the insight that, if the neurons' voltages are initially distributed according to a Lorentzian distribution, then the voltage distribution density continues to be Lorentzian at all times%
\footnote{%
    Recently, it was shown that for QIF neurons with Lorentzian heterogeneity, the voltage distribution density always converges to a Lorentzian independent from the initial distribution of voltages~\cite{pietras2023exact}. 
}.
The two defining parameters of the Lorentzian voltage distribution---mean and width---correspond directly to the mean voltage and the population firing rate, whose evolution is neatly described by the firing rate equations.
Lately, the Montbri\'o-Paz\'o-Roxin-approach has been extended to other forms of heterogeneity beyond Lorentzian distributed excitability (or inputs), see, e.g., \cite{klinshov2021reduction,pyragas2021dynamics,pyragas2022mean,pyragas2023effect}, but the resulting firing rate equations are higher dimensional, challenging to analyze, and their correspondence to the population firing rate and mean voltage is more involved.
Hence, what is so special about the Lorentzian distribution that simplifies the mean-field theory to large extent?
And, are the collective dynamics for Lorentzian heterogeneity qualitatively really different from the network behavior for other types of heterogeneity?

To make analytical progress---and to keep the theory as simple as possible while guaranteeing the generality of the results---, we study the activity of large heterogeneous populations of quadratic integrate-and-fire (QIF) neurons, whose
dynamics are given by
\begin{equation}
\begin{aligned}
    &\dot V_j = V_j^2 + \eta_j, \qquad \text{for } j = 1,2,\dots,\\
    &\text{if } V_j \to \infty, \text{ then neuron $j$ spikes and its voltage is reset to } V_j \leftarrow -\infty.
\end{aligned}\label{eq:QIF}
\end{equation}
The $\eta_{j=1,2,\dots}$ are excitability parameters, or inputs, drawn from a probability distribution $g_0(\eta)$ with mean $\bar\eta$ and unit half-width at half-maximum (HWHM).
The sign of $\eta_j$ determines whether the QIF neuron is either excitable ($\eta_j \le 0$) or oscillatory with frequency $f_j = \sqrt{\eta_j}/\pi$ if $\eta_j>0$.
Given a distribution $g_0(\eta)$ of excitability parameters, a population typically comprises excitable as well as oscillatory neurons. 
Furthermore, we assume that the mean of the distribution, i.e.~the mean input $\bar\eta$, can depend on the mean fields of the population.
The two most characteristic mean fields are the mean voltage $v(t)$ and the population firing rate $r(t)$,
\begin{subequations}
\begin{align}
    &v(t) = \langle V_j(t)\rangle,\\
    &r(t) = \lim_{\tau_r\to0} R_{\tau_r}(t) := \lim_{\tau_r\to0} \Big\langle {\sum_k} \frac{1}{\tau_r} \int_{t-\tau_r}^t \delta(\zeta-t_j^k) d\zeta \Big\rangle,
    \label{firing_rate_formula}
\end{align}
\end{subequations}
where $t_j^k$ is the $k$th spike time of neuron $j$, $\tau_r$ a time window of spike events, and $\langle (\cdot)_j \rangle$ denotes the population average; 
throughout this manuscript, we will use the integral-description, 
e.g., for the mean voltage we write $\langle V_j(t)\rangle=\int_{\mathbb R} V(t|\eta) g(\eta) d\eta$, to describe both finite ($N<\infty$) and infinite ($N\to\infty$) sums of the type $\langle V_j(t)\rangle = \frac{1}{N}\sum_{j=1}^N V_j(t)$, where we identify $V(t|\eta_j)=V_j(t)$ and $g(\eta)$ is the probability distribution of the excitability parameter $\eta_j$ with $j=1,2,\dots,N$.

\Cref{eq:QIF} is the canonical description for a heterogeneous population of QIF neurons assuming that they are globally (``all-to-all'') coupled via chemical synapses of strength $J\in \mathbb R$ and via electrical synapses of strength $g\ge 0$ and receive a common external input $I(t)$, 
\begin{subequations}\label{eq:QIF_all2all}
\begin{align}
    \tau_m \dot V_j &=\phantom{:} a V_j^2 + a\Delta \xi_j + I(t) + J \tau_m s(t) + g [v(t) - V_j],
\end{align}
where $\tau_m>0$ is the membrane time constant and $a>0$ a constant. 
Heterogeneity is characterized through individual inputs $\xi_j$ that follow a normalized distribution with zero mean and unit HWHM.
With the scaling parameter $\Delta>0$, the variance of the input across the population is then proportional to $a\Delta$.
Electrical coupling is mediated through the mean voltage $v(t)$, whereas chemical coupling is mediated through the mean field $s(t)$ that depends on the population firing rate $r(t)$, e.g, with first-order synaptic kinetics and decay time constant $\tau_s$,
\begin{align}
    &\tau_s \dot s = -s + r(t).
\end{align}
\end{subequations}

\begin{remark}
\Cref{eq:QIF} can be obtained from \cref{eq:QIF_all2all}
after nondimensionalization \cref{nondimensionalization}, rescaling of variables variables \cref{rescaling}, and shifting the voltage variable \cref{shifting} according to
\begin{subequations}\label{transformation}
    \begin{align}
        &\tilde I = I/(a\Delta), \quad \tilde J = J/(a \sqrt{\Delta}), \quad \tilde g = g/(a \sqrt{\Delta}), \quad \tilde \tau = a \sqrt{\Delta} \tau_s/\tau_m,
        \label{nondimensionalization}\\
        &\tilde t = a \sqrt{\Delta} t/\tau_m, \quad \tilde V_j = V_j/\sqrt{\Delta}, \quad \tilde r= \tau_m r/\sqrt{\Delta}, \quad \tilde v = v/\sqrt{\Delta}, \quad \tilde s = \tau_m s/\sqrt{\Delta},
        \label{rescaling}\\
        &\tilde U_j = (\tilde V_j - \tilde g/2), \label{shifting}
    \end{align}
\end{subequations}
which yields the QIF dynamics 
\begin{subequations}\label{eq:U_all}
\begin{align}
    \tilde U_j' &= \tilde U_j^2 + \tilde \eta_j, \label{eq:Udyn}\quad \text{with} \quad 
    \tilde \eta_j = \big[ \tilde I(\tilde t) - \tfrac{\tilde g}{4} + \tilde J \tilde s(\tilde t) + \tilde g \tilde v(\tilde t) \big] + \xi_j, \\
    \tilde s' &= (- \tilde s + \tilde r)/\tilde \tau;
\end{align}
\end{subequations}
here, $'$ denotes the derivative with respect to the rescaled time $\tilde t$.
The dynamics \cref{eq:Udyn} are identical to \cref{eq:QIF} as the $\tilde \eta_j$ are distributed with the same shape as the $\xi_j$, i.e.~the HFHM of the distribution is unity, and the mean of the $\tilde \eta_j$ is common to all neurons and depends explicitly on the mean fields, see the term in brackets in \cref{eq:Udyn}.
This exercise underscores the generality of \cref{eq:QIF} for studying 
heterogeneous populations of globally coupled QIF neurons.
\end{remark}

\begin{figure}[ht!]
    \centering{
    \includegraphics[width=7.5cm]{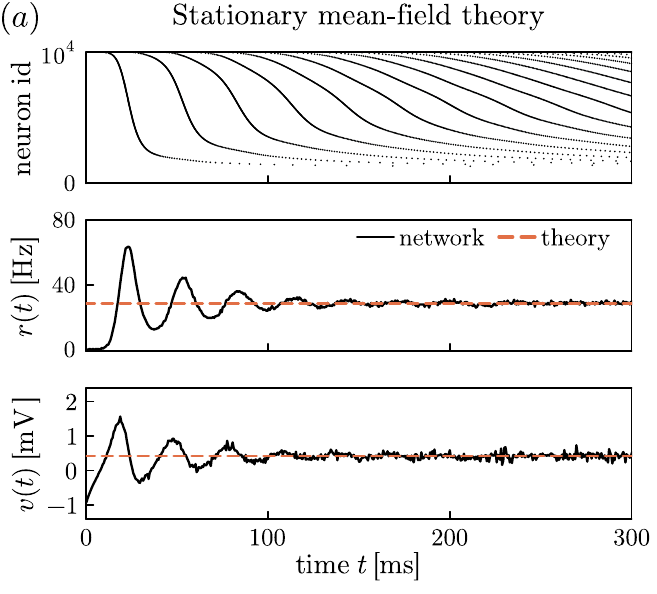}
    \hspace{0.1cm}
    \includegraphics[width=7.5cm]{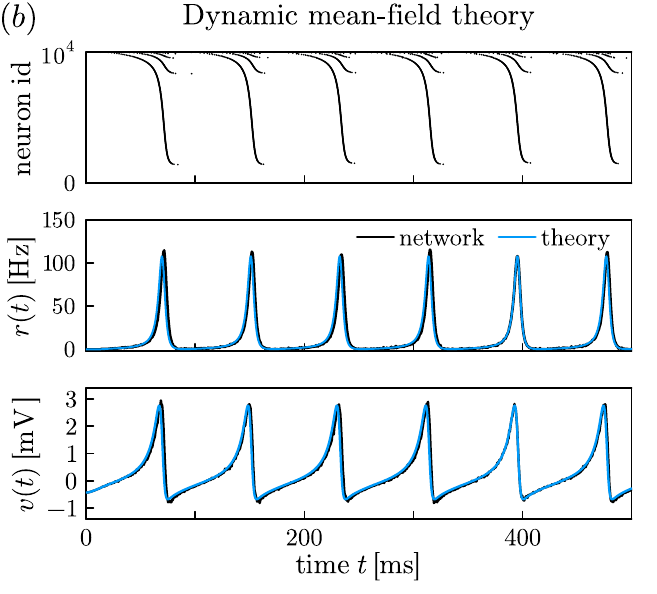}
    \caption{(a) Stationary and (b) dynamic mean-field theory for heterogeneous populations of globally coupled QIF neurons captures the stationary (red-dashed) and transient (blue-solid) collective dynamics, respectively.
    From top to bottom: raster plot of the neurons' spike times, population firing rate $r(t)$ and mean voltage $v(t)$.
    Network simulations were performed for $N=10000$ neurons with Gaussian heterogeneity following \cref{eq:QIF_all2all} with $\tau_m=10$ms $=2\tau_s$, $a=1,J=0$ and (a) $\Delta=2.1,I(t)=-0.2,g=3.0$ or (b) $\Delta=1,I(t)=0.8,g=1.0$,
    using an Euler-timestep $dt=5\times10^{-4}\tau_m$. 
    The firing rate $r(t)$ was obtained from \cref{firing_rate_formula} with $\tau_r=10^{-2}$ms and similarly the mean voltage $v(t)$.
    The corresponding mean-field theories are detailed in \cref{subsubsec:Gauss} for stationary solutions and in \cref{sec:nonuniversal} for dynamic solutions.
    In (a) we started the network simulation from random initial conditions at $t=-100$ms with $I(t)=-1.0$ and increased the mean input to $I(t>0)=-0.2$ at $t=0$. 
    In (b) we simulated the $q$-Gaussian collective dynamics \cref{eq:W_multi_all} with $n=100$ as an accurate proxy for Gaussian heterogeneity.
    We started the network as well as the mean fields from random initial conditions at $t=-100$ms with $\Delta=0.5$, we then increased $\Delta$ to $1$ at time $t=0$, leading to collective oscillations.}
    \label{fig:netsim}
    }
\end{figure}

The goal of this manuscript is to work out similarities and nonuniversal behavior of the collective dynamics of globally coupled QIF neurons with different types of unimodal heterogeneity. 
To this end, we distribute the excitability parameters $\eta_j$ across the population according to, e.g., Lorentzian, Gaussian or uniform distributions $g_0(\eta)$.
In anticipation of our results, nonuniversal behavior will become evident if the population consists of both excitable and self-oscillatory units, that is, when the mean $\bar\eta\approx 0$.
Recall that $\bar\eta$ can equivalently be interpreted as the mean input and hence depend on mean fields, such as the population firing rate or the mean voltage, that evolve in time.
In \cref{sec:SMF}, we deal with macroscopic stationary regimes (``stationary mean-field theory''; \cref{fig:netsim}a), where the mean fields of the population converge to stationary solutions;
in this case, $\bar\eta$ is constant in time and the population firing rate $r(t)=r_0$ and mean voltage $v(t)=v_0$ as well as the firing rate and voltage distributions can be computed self-consistently.
In \cref{sec:DMF}, we allow $\bar\eta= \bar\eta(t)$ to depend on the dynamics of the mean fields through recurrent chemical and electrical coupling
and present an analytically tractable ``dynamic mean-field theory'' (\cref{fig:netsim}b) for a broad class of heterogeneity, which includes Lorentzian distributions of the $\eta_j$ and approximates Gaussian as well as uniform distributions. 
In \cref{sec:nonuniversal}, we compare the collective dynamics for these different types of heterogeneity with particular focus on heterogeneity-induced and nonuniversal transitions to synchrony.
Conclusions will be drawn and discussed in \cref{sec:conclusion}.

\section{Stationary mean-field theory}
\label{sec:SMF}
Neurons typically receive various external and recurrent synaptic inputs that can vary in time, which makes analytic progress for the collective behavior of heterogeneous populations cumbersome. 
In macroscopic stationary states, however, it is possible to obtain exact solutions for the firing rate $r(t)=r_0$ and the mean membrane potential $v(t)=v_0$ of QIF neurons~\eqref{eq:QIF} when the mean input $\bar\eta(t) \equiv \bar\eta$ is constant in time.
The self-consistency arguments detailed below hold independent of the distribution $g_0(\eta)$; the subscript $0$ here refers to the distribution being normalized with unit HWHM, which we assume without loss of generality.
First, we will provide general expressions for the stationary firing rate $r_0$, mean voltage $v_0$ as well as the stationary firing rate and voltage distributions $P_0(f)$ and $P_0(V)$.
Then, in \cref{subsec:RandV_Ldistributions}, 
we will present a useful relation between the firing rate, mean voltage and an integral over the heterogeneity, which can be conveniently evaluated with Cauchy's Residue Theorem if the distribution $g_0(\eta)$ has a finite number of finite poles in the complex $\eta$-plane.
In particular, we will focus on rational and $q$-Gaussian distributions $Q_n(\eta)$ and $G_n(\eta)$ that converge to uniform and Gaussian distributions, respectively, in the limit $n\to \infty$. 
Moreover, for $n=1$, $Q_n(\eta)$ and $G_n(\eta)$ coincide and describe a Lorentzian distribution, which yields particularly simple expressions for the quantities $r_0,v_0, P_0(f)$ and $P_0(V)$, see \cref{subsec:RandV_Lorentzian}.
In anticipation of a ``dynamic mean-field theory'' put forward in \cref{sec:DMF}, we will consider the dynamics of infinitely many QIF neurons~\eqref{eq:QIF} in the continuum limit $N\to\infty$, where the heterogeneous inputs $\eta$ are drawn from a piecewise continuous distribution density $g_0(\eta)$ with mean $\bar\eta\in \mathbb R$. We note, however, that the results in this~\cref{sec:SMF} equally hold for heterogeneous populations of finitely many QIF neurons; in this case, the integrals over $g_0(\eta)$ should be interpreted as discrete sums.

\subsection{Population firing rate and firing rate distribution}
\label{subsec:firing}
To begin, we consider one of the most characteristic mean fields of neuronal networks, the \textbf{population firing rate} \cref{firing_rate_formula}, which is formally defined as the fraction of neurons that fire a spike within an infinitesimally small time window.
The stationary firing rate $r(t)=r_0$ of QIF neurons~\eqref{eq:QIF}
can also be determined self-consistently, namely from the individual steady-state frequencies, or firing rates,
\begin{equation}\label{eq:individual_frequencies}
    f_j = f(\eta_j) = \sqrt{\eta_j}/\pi \times \Theta(\eta_j) >0,
\end{equation}
where $\Theta$ denotes the Heaviside-function.
Then, the population firing rate is obtained by averaging the individual firing rates $f_j$ over the population,
\begin{equation}
    r_0 = \int_\mathbb{R} f(\eta) g_0(\eta)d\eta = \frac1\pi\int_0^\infty \sqrt{\eta} g_0(\eta) d\eta \ .
\end{equation}
Alternatively, one can obtain $r_0$ as the mean of the individual firing rates $f_j$,
\begin{align*}
    r_0 = \int_0^\infty f P_0(f) df,
\end{align*}
where $P_0(f)$ is the {\bf firing rate distribution}.
We can determine $P_0(f)$ from the relation~\eqref{eq:individual_frequencies} between the firing rate $f_j$ and the excitability parameter $\eta_j$.
Formally, we have
\begin{equation}\label{eq:firing_rate_dist}
    P_0(f) = \int_0^\infty \delta\big(f - f(\eta)\big) g_0(\eta) d\eta = 2\pi^2 f g_0( \pi^2f^2 ) \times \Theta(f),
\end{equation}
where we used the transform of variables for Dirac $\delta$ functions, i.e.\ $\delta(h(x)) = | h'(x^*)|^{-1} \delta(x-x^*)$ where $h(x^*)=0$.
From the firing rate distribution~\eqref{eq:firing_rate_dist}, we can compute the population firing rate (with inverse transform $\eta_j = \pi^2f_j^2$)
\begin{equation}
    \langle f \rangle = \int_0^\infty fP_0(f)df = 2 \int_0^\infty \pi^2 f^2 g_0( \pi^2f^2 ) df = \frac{1}{\pi} \int_0^\infty \sqrt{\eta}g_0(\eta)d\eta = r_0.
\end{equation}

\subsection{Mean membrane potential and voltage distribution}
\label{subsec:voltage}
Next to the firing rate $r(t)$, another characteristic mean field of neuronal networks is the mean voltage, or mean membrane potential, $v(t)=\langle V_j\rangle$.
To compute the stationary mean voltage $v(t)=v_0$, we need to know the asymptotic mean voltages $\bar V_j = \langle V_j(t) \rangle_t$ of the individual neurons; here we take the average over time. 
Periodically firing QIF neurons with $\eta_j>0$ have zero average membrane potential $\bar V_j = 0$ because of the symmetry of \cref{eq:QIF}, and hence do not contribute to the mean voltage $v_0$.
Excitable QIF neurons with $\eta_j<0$, by contrast, will converge to their asymptotic voltage 
\begin{equation}
    \bar V_j = \bar V(\eta_j) = -\sqrt{-\eta_j} \times \Theta(-\eta_j) < 0.
    \label{eq:asymptotic_voltage_excitable}
\end{equation}
Thus, the mean voltage can be found by solving
\begin{equation}
    v_0 = \int_\mathbb{R} \bar V(\eta) g_0(\eta)d\eta = - \int_{-\infty}^0 \sqrt{-\eta} g_0(\eta) d\eta.
    \label{eq:v0_first}
\end{equation}

The stationary mean voltage $v_0$ can also be obtained as the mean of the distribution of membrane potentials, or {\bf voltage distribution}, $P_0(V)$. 
To compute $P_0(V)$, we capitalize on the convenience that we can assign asymptotic conditional voltage distributions for the QIF dynamics~\eqref{eq:QIF}. 
That is, after initial transients at any time $t$ large enough, the probability $\rho(V|\eta)$ that a neuron with excitability parameter $\eta$ has voltage $V$ is given by
\begin{equation}\label{eq:conditional_voltage_distribution}
    \rho(V | \eta) = \frac{f(\eta)}{V^2 + \eta} \times \Theta(\eta) + \delta(V + \sqrt{-\eta}) \times \Theta(-\eta).
\end{equation}
The first term in \cref{eq:conditional_voltage_distribution} describes the voltage distribution density for an oscillatory neuron ($\eta>0$), which is inversely proportional to its speed and $f(\eta)$ given by \cref{eq:individual_frequencies} turns out to be the normalization constant~\cite{kuramoto1984chemical}.
The second term in \cref{eq:conditional_voltage_distribution} corresponds to the asymptotic voltage $\bar V(\eta)$ of an excitable neuron ($\eta\leq0$) described by \cref{eq:asymptotic_voltage_excitable}.
Averaging \cref{eq:conditional_voltage_distribution} both over $\eta$ and over the individual membrane potentials $V\in \mathbb{R}$, yields the mean voltage
\begin{equation}
    \iint_\mathbb{R} \rho(V|\eta) g_0(\eta) d\eta dV = \int_\mathbb{R} \int_{-\infty}^0 V\delta(V+\sqrt{-\eta})g_0(\eta)d\eta dV= -\int_{-\infty}^0 \sqrt{-\eta} g_0(\eta)d\eta=v_0, 
    \label{eq:v0_from_meanconditional}
\end{equation}
where we used that the first term in \cref{eq:conditional_voltage_distribution} is symmetric with respect to $V=0$ and does not contribute to the mean $v_0$---in line with \cref{eq:v0_first}.
Averaging the asymptotic conditional voltage distribution \cref{eq:conditional_voltage_distribution} over the heterogeneity parameter $\eta$, yields the stationary total voltage density $P(V,t) = P_0(V)$ 
\begin{equation}\label{eq:total_voltage_P0V}
    P_0(V) = \int_\mathbb{R} \rho(V|\eta) g_0(\eta) d\eta = \frac1\pi \int_0^\infty \frac{\sqrt{\eta} \ g_0(\eta)}{V^2 + \eta} d\eta + \int_{-\infty}^0 \delta(V+\sqrt{-\eta})g_0(\eta)d\eta.
\end{equation}
The first integral on the right-hand side of \cref{eq:total_voltage_P0V} will contribute a term that is symmetric about $V=0$. 
The second integral, by contrast, only contributes to $P_0(V)$ for negative $V$.
We can simplify the integral by applying the change of variable transform for Dirac $\delta$ functions.
That is, with $\delta(h(x)) = | h'(x^*)|^{-1} \delta(x-x^*)$ where $h(x^*)=0$ with $h(x) = V + \sqrt{-x}$ and $h'(x) = -1/ (2\sqrt{-x})$, we eventually obtain
\begin{equation}\label{eq:total_voltage_2nd_integral}
    \int_{-\infty}^0 \delta(V+\sqrt{-\eta})g_0(\eta)d\eta = 2 |V| g_0(-V^2) \times \Theta(-V).
\end{equation}
We can use \cref{eq:total_voltage_2nd_integral} to compute the mean voltage $v_0$, alternatively to \cref{eq:v0_first,eq:v0_from_meanconditional},
as
\begin{equation}
    \langle V \rangle = \int_\mathbb{R} V P_0(V) dV = \int_{-\infty}^0 -2V^2g_0(-V^2)dV = - \int_{-\infty}^0 \sqrt{-\eta}g_0(\eta)d\eta = v_0,
    \label{eq:v0_integral}
\end{equation}
where we used the fixed point relation $V = -\sqrt{-\eta} \Leftrightarrow \eta = -V^2$ for the change of variables between the integrals.

\subsection{Firing rate and voltage distributions of heterogeneous populations}
\begin{figure}[ht!]
    \centering{
        \includegraphics[width=15cm]{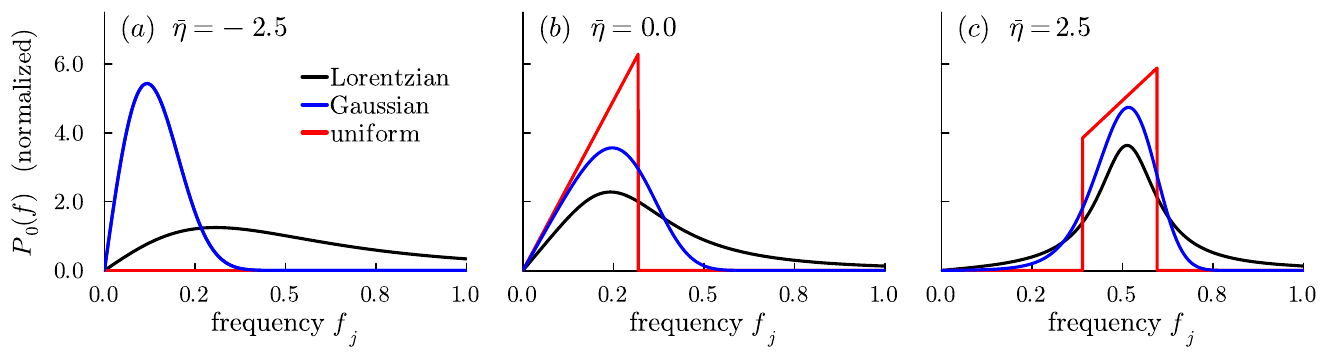}
        \caption{Firing rate distributions $P_0(f)$ for heterogeneous populations of QIF neurons subject to Lorentzian, uniform and Gaussian distributed inputs with mean (a) $\bar\eta= -2.5$, (b) $\bar\eta= 0$, (c) $\bar\eta= 2.5$ and unit HWHM.
        The $P_0(f)$ are normalized with respect to the fraction of neurons that are periodically firing.
        In $(a)$, uniform heterogeneity suppresses firing of all neurons for strong negative inputs. 
        The fat tails of Lorentzian distributions make the firing frequencies $f_j$ more widespread compared to Gaussian heterogeneity.
        This effect, however, vanishes for larger mean inputs as the firing rate distributions become narrower and peak about the mean firing rate, cf.~panel $(c)$.
        }
    }
    \label{fig:fr_dist}
\end{figure}

The expressions \cref{eq:firing_rate_dist,eq:total_voltage_P0V} for the firing rate distribution $P_0(f)$ and the voltage distribution $P_0(V)$ are general and can be applied for heterogeneous populations of QIF neurons with arbitrary input distributions $g_0(\eta)$. 
For instance, in \cref{fig:fr_dist}, we present firing rate distributions $P_0(f)$ for the typical unimodal distributions---Lorentzian (black), Gaussian (blue) and uniform (red)---for three different means (a) $\bar\eta= -2.5$, (b) $\bar\eta= 0$, (c) $\bar\eta= 2.5$ with unit HWHM; see the following \cref{subsec:RandV_Ldistributions} for the corresponding formulae and for sketches of the distributions.
For large negative inputs such that the HWHM $\Delta$ is less than the (absolute value of the) mean, $\Delta < -\bar\eta$, see panel $(a)$, the compact support of the uniform distribution yields that all the neurons are quiescent and $P_0(f)$ coincides with the x-axis.
By contrast, Lorentzian and Gaussian distributions have infinite support, so that even for negative mean inputs there is a considerable fraction of neurons that are firing with frequencies $f_j>0$. 
The fat tails of the Lorentzian, however, lead to a wide spread of the firing frequencies $f_j$, whereas the firing rate distribution for Gaussian heterogeneity is narrowed about its peak.
With increasing mean inputs, see panels ($b$) and ($c$), the firing rate distributions for Gaussians and Lorentzians approximate each other and also the firing rate distribution for uniform heterogeneity becomes centered around the peaks of the other two.

\begin{figure}[htb!]
    \centering{
        \includegraphics[width=15cm]{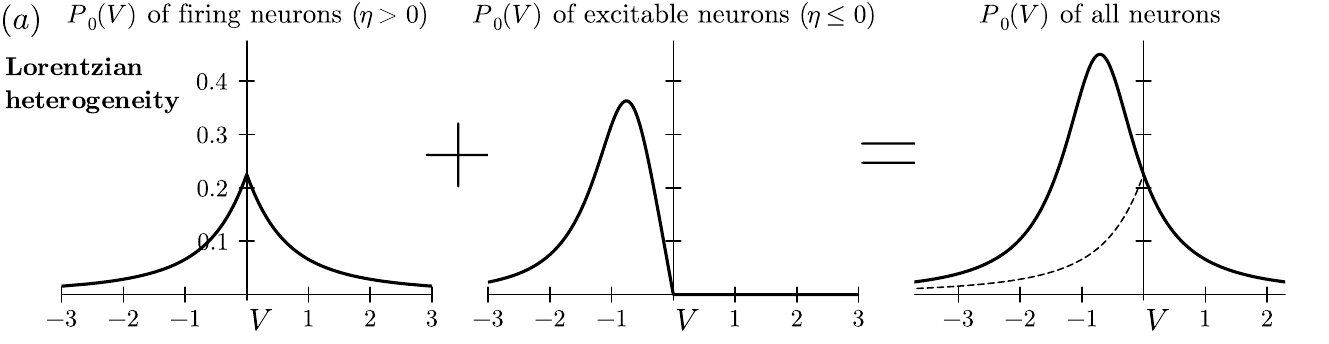}\\[1em]
        \includegraphics[width=15cm]{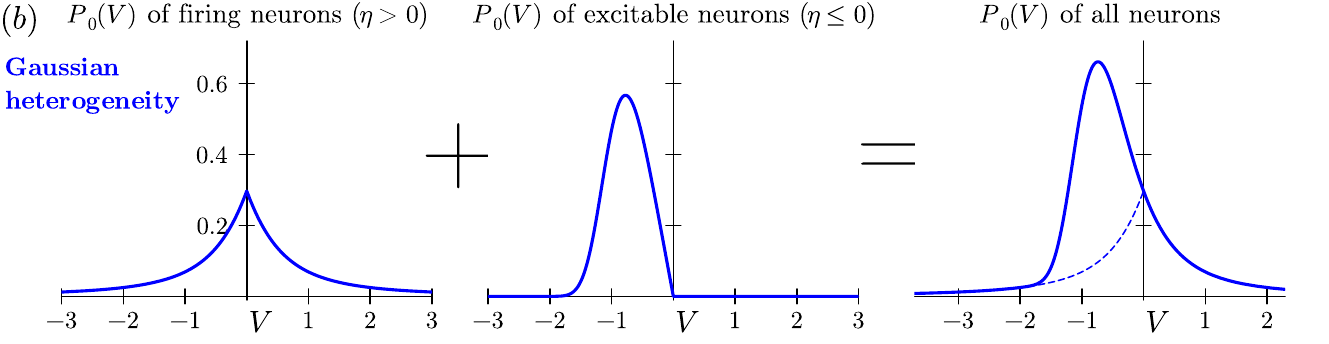}\\[1em]
        \includegraphics[width=15cm]{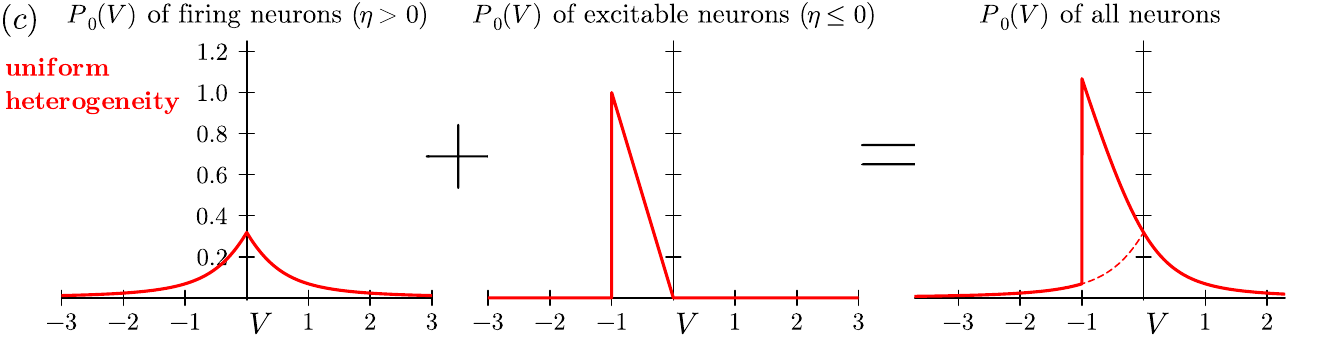}\\
        \caption{Total voltage distributions $P_0(V)$ in macroscopic stationary states given by \cref{eq:total_voltage_P0V} for (a) Lorentzian, (b) Gaussian, and (c) uniform heterogeneity.
        The voltage distribution density of periodically firing neurons is inversely proportional to their frequency, so that their contribution to $P_0(V)$ is symmetric with respect to $V=0$ (left panel).
        Excitable neurons settle in their fixed points, so that they contribute to $P_0(V)$ only for $V\leq 0$ (middle panel).
        The sum of the two gives the total voltage distribution (right panel), which in the case of Lorentzian heterogeneity turns out to be a Lorentzian distribution (\cref{subsec:RandV_Ldistributions}); for other types of heterogeneity, the resulting voltage distribution does not correspond to the distribution of inputs and can be skewed or discontinuous.
        The HWHM for all three heterogeneities is $\Delta=1$ and their mean is $\bar\eta=0$, so that half of the population is firing and half excitable.
        Nonetheless, the voltage distributions are notably different: The fat tails of the Lorentzian manifest through a larger spread of voltages, whereas uniform heterogeneity leads to more narrow voltage distributions that will favor synchrony and collective oscillations, see also \cref{sec:nonuniversal}.}
    }
    \label{fig:v_dist}
\end{figure}

In \cref{fig:v_dist}, we present voltage distributions $P_0(V)$ for Lorentzian (black), Gaussian (blue) and uniform (red) heterogeneity, with zero mean and unit HWHM, but the results are generic for any mean input $\bar\eta \in \mathbb R$.
According to \cref{eq:total_voltage_P0V}, it is convenient to consider the asymptotic voltage distributions separately for those neurons that are periodically firing, or self-oscillatory (left panels), and for those that are quiescent (middle panels).
The voltage distribution $P_0(V)$ of firing neurons is clearly symmetric about $V=0$, whereas $P_0(V)$ of quiescent neurons is only defined for $V<0$. 
The sum of the two parts results, in general, in a skewed total voltage distribution (right panel), cf.~\cref{fig:v_dist}(b,c) for Gaussian and uniform heterogeneity.
Curiously, however, in case of Lorentzian heterogeneity~(\cref{fig:v_dist}a), the shape of the total voltage distribution $P_0(V)$ seems again Lorentzian---we will come back to this observation in \cref{subsubsec:Lorentzian_P0V}.
As in the case of the firing rate distributions, the fat tails of the Lorentzian manifest through a larger spread of voltages compared to the narrower $P_0(V)$ for Gaussian and uniform heterogeneity.
Narrower voltage distributions may favor the emergence of synchronization, e.g., collective oscillations; we will investigate the consequences for collective dynamics in more detail in \cref{sec:nonuniversal}.

As we have seen in \cref{subsec:firing,subsec:voltage}, we can use the firing rate and voltage distributions also to determine the population firing rate $r_0$ and the mean voltage $v_0$. 
In the following \cref{subsec:RandV_Ldistributions}, we will propose a neat mathematical trick to  compute $r_0$ and $v_0$ directly for analytically favorable distributions $g_0(\eta)$.

\subsection{Firing rate and mean voltage via residue theorem}
\label{subsec:RandV_Ldistributions}
The stationary population firing rate $r_0$ and the mean membrane potential $v_0$ for arbitrary heterogeneous populations of QIF neurons~\eqref{eq:QIF}, determined via the distribution $g(\eta)$ of inputs, are
\begin{align}\label{eq:r0v0_means}
    r_0 = \frac1\pi\int_0^\infty \sqrt{\eta} g_0(\eta) d\eta 
    \quad \text{and} \quad
    v_0 = - \int_{-\infty}^0 \sqrt{-\eta} g_0(\eta) d\eta.
\end{align}
Note that if $g_0(\eta)$ is non-vanishing only for $\eta>0$, the mean voltage $v_0$ is equal to zero. 
Likewise, if $g_0(\eta)$ is non-zero only for $\eta<0$, then the population firing rate $r_0$ vanishes. 

\begin{figure}[ht!]
    \centering{
    \includegraphics[width=7.5cm]{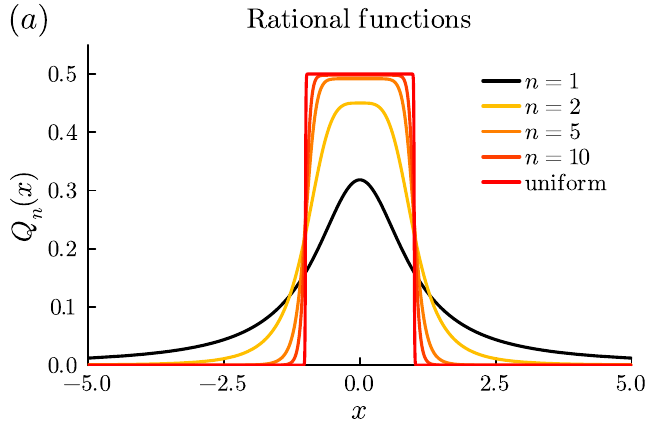}
    \includegraphics[width=7.5cm]{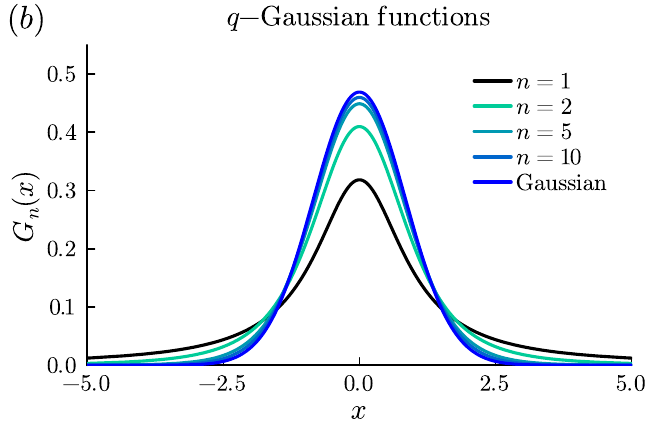}
    \caption{Rational and $q$-Gaussian distribution functions are admissible to the computation of the firing rate $r_0$ and the mean voltage $v_0$ via \cref{eq11} using Cauchy's Residue Theorem.
    (a) Rational distributions $Q_n$ interpolate between Lorentzian ($n=1$) and uniform ($n\to\infty$) distributions.
    (b) $q$-Gaussian distributions $G_n$ interpolate between Lorentzian ($n=1$) and Gaussian ($n\to\infty$) distributions. The Lorentzian has fatter tails than the other distributions. 
    }
    \label{fig:1}
    }
\end{figure}

Combining \cref{eq:r0v0_means} yields the useful relation
\begin{equation}
    \pi r_0 - iv_0 = \int_\mathbb{R} \sqrt{\eta} g_0(\eta)d\eta \ ,
    \label{eq10}
\end{equation}
where the integral on the right-hand side is defined along the real axis.
For certain distributions $g_0(\eta)$ of the heterogeneity parameter $\eta$, 
\cref{eq10} can conveniently be solved by applying Cauchy's Residue Theorem:
one closes the integral along the real line with an arc $\pm|\eta|\exp(i \vartheta)$ either in the upper ($+$) or the lower ($-$) complex half-plane with $|\eta| \to \infty$ and $\vartheta \in (-\pi,0)$. 
If this contour encloses finitely many poles, one can then evaluate the integral by computing the residues of these poles. 
As physically meaningful firing rates cannot become negative, i.e.~we require $\pi r_0 \ge 0$, we advise caution when choosing the poles in the upper or lower half-plane; one has to evaluate the resulting complex roots of the poles consequently either as primary or secondary solutions. 
It turns out that computations are straightforward for continuation of $\eta$ in the lower complex half-plane. 
Then, \cref{eq10} can be evaluated as
\begin{equation}
    \pi r_0 - iv_0 = -2\pi i \sum_k \mathrm{Res} (\sqrt{\eta}g_0(\eta); \zeta_k) \ ,
    \label{eq11}
\end{equation}
where the sum runs over the poles $\eta=\zeta_k$ of $g_0(\eta)$ in the lower half of the complex $\eta$-plane (that is the reason for the minus-sign).
Suitable distribution densities $g_0(\eta)$ have a finite number of finite poles, such as the (Cauchy-)Lorentzian distribution~\cite{ott2008low}, a superposition of multiple Lorentzians~\cite{martens2009exact,pazo2009existence,omel2012nonuniversal,pyragas2021dynamics}, or the general family of distributions $L_n^m(\eta)$, for $n>0, mn>1$, that was proposed in~\cite{lafuerza2010nonuniversal},
\begin{equation}
    L_n^m(\eta) = \frac{n \Gamma(m)}{2\Gamma(m-1/n)\Gamma(1/n)} \frac{d^{nm-1}}{(|\eta-\bar\eta|^n + d^n)^m}.
    \label{eq:Lnm}
\end{equation}
The variance of $L_n^m$ is finite only for $mn>3$ and it is given by $\sigma^2=d^2 \frac{\Gamma(m-3/n)\Gamma(3/n)}{\Gamma(m-1/n)\Gamma(1/n)}$.
The Lorentzian distribution corresponds to $L_2^1(\eta)$ and has, hence, infinite variance.
Previous studies have used \cref{eq:Lnm} to interpolate between a Lorentzian and a compact uniform distribution~\cite{skardal2018low,pietras2018first,pyragas2023effect} via so-called rational distributions, or between a Lorentzian and Gaussian distribution via so-called $q$-Gaussian distributions~\cite{pyragas2022mean}; see \cref{fig:1}.
In the following, we present how \cref{eq11} simplifies for each of these distributions.

\subsubsection{Rational distributions}
From \cref{eq:Lnm}, we obtain rational distributions as $Q_n(\eta) := L^1_{2n}(\eta)$ with parameter $n=1,2,\dots$ and $d=1$, which become a uniform distribution in the limit $n\to\infty$. 
The $Q_n(\eta)$ read
\begin{equation}
    Q_n(\eta) = \frac{n}\pi \sin\left(\frac\pi{2n}\right) \frac{1}{(\eta-\bar\eta)^{2n} + 1},
\end{equation}
with mean $\bar\eta$ and unit half-width-at-half-maximum (HWHM) for all $n$.
$Q_n(\eta)$ has $n$ simple poles $\zeta_k$ in the lower half-plane and $n$ complex conjugate poles $\zeta_k^*$ in the upper half-plane,
\begin{equation}
    \zeta_k = \bar\eta + \alpha_k, \quad \text{with} \quad \alpha_k = \exp\left[ -\frac{i\pi}{2n} (2k-1) \right], \quad k=1,\dots,n.
    \label{eq:zetak_rational}
\end{equation}
For rational distributions $Q_n(\eta)$, \cref{eq11} yields
\begin{equation}
    \pi r_0 - iv_0 = \int_\mathbb{R} \sqrt{\eta}Q_n(\eta)d\eta = i \sin\Big( \frac\pi{2n} \Big) \sum_{k=1}^n \alpha_k \sqrt{\zeta_k} \ .
    \label{eq:rv0_Qn}
\end{equation}
In general, the stationary mean firing rate $r_0$ and membrane potential $v_0$ have to be computed from the sum of $n$ complex square roots.

\begin{proof}[Proof of \cref{eq:rv0_Qn}]
The residue of $Q_n(\eta)$ at a pole $\eta=\zeta_k = \bar\eta + \alpha_k$ with $\bar\eta \in \mathbb R$ and $\alpha_k$ given by \cref{eq:zetak_rational} can be computed as 
\begin{align*}
    \mathrm{Res}(Q_n,\zeta_k) &= \lim_{\eta \to \zeta_k} (\eta-\zeta_k) Q_n(\eta) = \frac{n}{\pi} \sin\Big(\frac{\pi}{2n}\Big) \lim_{\alpha \to \alpha_k}  \frac{ (\alpha - \alpha_k)}{\alpha^{2n} +1} \\
    &= \frac{n}{\pi} \sin\Big(\frac{\pi}{2n}\Big) (\alpha_k-\alpha_{-k})^{-1}\prod_{j=1,j\neq k}^n (\alpha_k - \alpha_j)^{-1}(\alpha_k - \alpha_{-j})^{-1},
\end{align*}
where we used that the $\alpha_k, \alpha_{-k}=\alpha^*_k$ with $k=1,\dots,n$, are the $2n$-th roots of minus unity, see \cref{eq:zetak_rational}. That is why, in general, it holds
\begin{align*}
    &f(x) := x^{2n} + 1 = \prod_{j=1}^{n} (x-\alpha_j)(x-\alpha_{-j})\\
    &\Longrightarrow \quad
    f'(x) = 2n x^{2n-1} = \sum_{l=1}^{n} \big[ (x-\alpha_l) + (x-\alpha_{-l})\big] \prod_{j=1,j\neq l}^{n} (x-\alpha_j)(x-\alpha_{-j}).
\end{align*}
When evaluating $f'(x)$ at $x=\alpha_k$, then all but one of the summands vanish, so
\begin{align*}
    f'(\alpha_k) = (\alpha_k-\alpha_{-k}) \prod_{j=1,j\neq k}^{n} (x-\alpha_j)(x-\alpha_{-j})= 2n \alpha_k^{2n-1} = \frac{-2n}{\alpha_k},
\end{align*}
where we used that $\alpha_k^{2n}=-1$.
In sum, we obtain
\begin{align}
    \mathrm{Res}(Q_n,\zeta_k) = \frac{-1}{2\pi} \sin\Big(\frac{\pi}{2n}\Big) \alpha_k.
    \label{eq:lemma_residue_rational}
\end{align}
\end{proof}

\subsubsection{Uniform distribution}
In the limit $n\to\infty$, $Q_n(\eta)$ converges to a compact uniform distribution of half width $1$ centered at $\bar\eta$, i.e., 
\begin{align}
    Q_\infty(\eta) = \begin{cases}
        \frac{1}{2}, \quad &\text{if} \quad \eta \in [q_\text{min},q_\text{max}]:=[\bar\eta-1,\bar\eta+1],\\
        0, &\text{otherwise},
    \end{cases}
\end{align}
where $q_{\text{max/min}} = \bar\eta \pm 1$.
Then, $r_0$ and $v_0$ simplify as
\begin{subequations}
\label{eq:rv0_uniform}
\begin{alignat}{2}
    &\text{if } \bar\eta + 1 \le 0: \quad &&r_0 = 0 \quad \text{and}\quad v_0 = -\frac{(|\bar\eta| + 1)^{3/2}-(|\bar\eta| - 1)^{3/2}}{3},\\    
    &\text{if } \bar\eta - 1 > 0:  &&r_0 = \frac{(\bar\eta + 1)^{3/2}-(\bar\eta - 1)^{3/2}}{3\pi} \quad \text{and} \quad v_0 = 0,\\
    &\text{if } |\bar\eta| < 1:  &&r_0 = \frac{(\bar\eta + 1)^{3/2}}{3\pi} \quad \text{and} \quad v_0 =  -\frac{(|\bar\eta| + 1)^{3/2}}{3} \ ;
\end{alignat}
note that in the last case, $|\bar\eta| < 1$, zero lies within the support of the uniform distribution, i.e.\ $ q_\text{min} < 0 < q_\text{max}$, and one has to split the integral when computing $r_0$ and $v_0$.
For $\bar\eta = 0$, we have $v_0=-\pi r_0$; this relation also holds for Gaussian and Lorentzian distributions, see below.
\end{subequations}

\subsubsection{$q$-Gaussian distribution}
The $q$-Gaussian distribution $G_n(\eta)= L_2^n(\eta)$, which is related to Student's $t$-distribution\footnote{The location-scale $t$-distribution $lst(\eta;\mu,\tau^2,\nu)$ becomes $G_n(\eta)$ with $\mu=\bar\eta$, $\nu=2n-1$ and $\tau^2= d^2/\nu$.}, becomes a Gaussian in the limit $n\to\infty$. 
Choosing $d = \beta_n^{-1/2}$ with $\beta_n=2^{1/n}-1$ ensures that $G_n(\eta)$ has unit HWHM for all $n=1,2,\dots$; consequently, the variance of the Gaussian distribution $G_\infty(\eta)$ is $\sigma^2=1/[2\log(2)]$.
From \cref{eq:Lnm}, $G_n(\eta)$ then reduces as~\cite{pyragas2022mean}
\begin{align}
    G_n(\eta) &= \frac{1}{\sqrt{\pi}}\frac{\Gamma(n) \sqrt{\beta_n}}{\Gamma(n-1/2) } \left[ 1 + \beta_n (\eta-\bar\eta)^2 \right]^{-n}\\
    &=\frac{1}{\sqrt{\pi}} \frac{\Gamma(n) }{\Gamma(n-1/2)} {\beta_n}^{(1-2n)/2} \left| \eta - \bar\eta + i \beta_n^{-1/2} \right|^{-2n}
    \nonumber
\end{align}
and has two poles of order $n$,
\begin{equation}
    \zeta_\pm = \bar\eta \pm i \beta_n^{-1/2} .
\end{equation}
To evaluate the right-hand side of \cref{eq11}, one now has to employ Cauchy's Residue Theorem for higher-order poles, which yields
\begin{equation}
    \pi r_0 - iv_0 = -\frac{2\sqrt{\pi} i}{(n-1)!} \frac{\Gamma(n) }{\Gamma(n-1/2)} \sqrt{\beta_n}^{1-2n} \lim_{\eta \to \zeta_-} \left[ \frac{\partial^{n-1}}{\partial \eta^{n-1}} \frac{\sqrt{\eta}}{(\eta - \zeta_+)^n}\right]  \ .
    \label{eq:rv0_Gn}
\end{equation}
Using the analytical tricks as detailed in~\cite{pyragas2022mean}, 
one can simplify \cref{eq:rv0_Gn} as 
\begin{subequations}\label{eq:r0v0_Gn_bk}
\begin{align}
    \pi r_0 - iv_0 = \int_\mathbb{R} \sqrt{\eta}G_n(\eta)d\eta = \sum_{k=1}^n b_k g_k,
\end{align}
where the $b_k\in \mathbb R$ are the same as in~\cite{pyragas2022mean}, cf.~their Eqs.(28-30), 
\begin{align}
    b_k = \frac{\Gamma(n-k/2)}{\Gamma(n-1/2)}\frac{\Gamma\big(n-(k-1)/2\big)}{\Gamma(n-k+1)}, \quad k = 1,\dots,n,
\end{align}
and they satisfy the recurrence relation
\begin{align}
    b_1 = 1,\quad b_k = \frac{n+1-k}{n-k/2} b_{k-1} \quad \text{for}\quad k=2,\dots,n.
\end{align}
On the other hand, the $g_k$ are
\begin{align}
    g_k = \frac{(i/\sqrt{\beta_n})^{k-1}}{(k-1)!} \lim_{\eta\to\zeta_-} \left[ \frac{\partial^{k-1}}{\partial \eta^{k-1}} \sqrt{\eta} \right] = \frac{(i/\sqrt{\beta_n})^{k-1}}{(k-1)!} \frac{\Gamma(3/2)}{\Gamma(5/2-k)} \zeta_-^{3/2-k},
\end{align}
for which we find the recurrence relation
\begin{align}
    g_1 = \sqrt{\zeta_-}, \quad g_k = \frac{i}{2\sqrt{\beta_n}} \frac{5-2k}{k-1} \frac{g_{k-1}}{\zeta_-} \quad \text{for} \quad k=2,\dots,n.
\end{align}
\end{subequations}
We can simplify the foregoing expressions \cref{eq:r0v0_Gn_bk} by combining $b_k$ and $g_k$ to obtain
\begin{subequations}\label{eq:r0v0_Gn}
    \begin{align}
    \pi r_0 - iv_0 = \int_\mathbb{R} \sqrt{\eta}G_n(\eta)d\eta = \sum_{k=1}^n \tilde \gamma_k (i /\sqrt{\beta_n})^{k-1} \zeta_-^{3/2-k} = \sum_{k=1}^n \gamma_{k}\zeta_-^{3/2-k},
    \end{align}
where the $\tilde \gamma_k = [\Gamma(3/2)\Gamma(n-k/2)\Gamma(n-k/2+1/2)]/[\Gamma(k)\Gamma(5/2-k)\Gamma(n-1/2)\Gamma(n-k+1)]\in \mathbb R$ satisfy the recurrence relation
\begin{align}
    \tilde\gamma_1 = 1,\quad \tilde\gamma_2=\frac12, \quad \tilde\gamma_k = \frac{(5-2k)(n+1-k)}{(k-1)(2n-k)} \tilde\gamma_{k-1} \quad \text{for}\quad k=3,\dots,n,
\end{align}
or alternatively the $\gamma_{k} = \tilde\gamma_k(i/\sqrt{\beta_n})^{k-1}\in \mathbb C$ satisfy the recurrence relation
\begin{align}
    \gamma_{1} = 1,
    \quad \gamma_{k} = \frac{i}{\sqrt{\beta_n}} \frac{(5-2k)(n+1-k)}{(k-1)(2n-k)} \gamma_{k-1} \quad \text{for}\quad k=2,\dots,n.
\end{align}
\end{subequations}

\subsubsection{Gaussian distribution}
\label{subsubsec:Gauss}
In the limit $n\to\infty$, the $q$-Gaussian distribution $G_n$ becomes the standard Gaussian
\begin{equation}
    G_\infty(\eta) = \frac{1}{\sqrt{2\pi\sigma^2}}e^{-\frac{(\eta-\bar\eta)^2}{2\sigma^2}},
\end{equation}
so that we can evaluate \cref{eq11} according to
\begin{gather}
    \pi r_0 - iv_0 = \int_\mathbb{R} \sqrt{\eta}G_\infty(\eta)d\eta \\
    \hspace{-1.5cm}= \frac{(1+i)e^{-\xi}}{4 \sqrt{2 \sigma^2}} \sqrt{\frac{\pi}{|\bar\eta|}}
    \label{eq:rv0_Gauss}
    \begin{cases}
        2\sigma^2 I_{1/4}(\xi) + \bar\eta^2 \left[I_{1/4}(\xi) - i I_{-1/4}(\xi) - i I_{3/4}(\xi)+I_{5/4}(\xi)\right], \; &\text{if } \bar\eta \ge 0,\\
        2\sigma^2 I_{1/4}(\xi) + \bar\eta^2 \left[I_{1/4}(\xi) + i I_{-1/4}(\xi) + i I_{3/4}(\xi)+I_{5/4}(\xi)\right], &\text{if } \bar\eta < 0,
    \end{cases}
    \nonumber
\end{gather}
where we abbreviated $\xi = \bar\eta^2/ (4 \sigma^2)$ and $I_\nu(x)$ denotes the modified Bessel function of first kind.
The Gaussian distribution has unit HWHM if $2\sigma^2=1/\log(2)$.
Either from \cref{eq:rv0_Gauss} or alternatively solving the following integrals, we find the stationary population firing rate for Gaussian heterogeneity
\begin{align}\label{eq:r0_Gauss}
    &r_0 = \frac{1}{\pi} \int_0^\infty \sqrt{\eta}G_\infty(\eta)d\eta = 2 \int_0^\infty (\pi f)^2 G_\infty( \pi^2 f^2 ) df 
    =\\ 
    &\hspace{-0.85cm}= \frac{e^{-\xi}}{4 \sqrt{2 \pi \sigma^2 |\bar\eta| }} 
    \begin{cases}
        2\sigma^2 I_{1/4}(\xi) + \bar\eta^2 \left[I_{1/4}(\xi) +I_{-1/4}(\xi)+I_{3/4}(\xi)+I_{5/4}(\xi)\right], \quad &\text{if } \bar\eta > 0,\\
        2\sigma^2 I_{1/4}(\xi) + \bar\eta^2 \left[ I_{1/4}(\xi) -I_{-1/4}(\xi)-I_{3/4}(\xi)+I_{5/4}(\xi)\right], &\text{if } \bar\eta < 0,
    \end{cases}\nonumber
\end{align}
and $r_0=\Gamma(3/4) [2^3 \pi^6 \log(2) ]^{-1/4}\approx 0.14341$ if $\bar\eta = 0$. 
The stationary mean voltage reads
\begin{gather}\label{eq:v0_Gauss}
    v_0 = -\int_{-\infty}^0 \sqrt{-\eta}G_\infty(\eta)d\eta \stackrel{\eqref{eq:v0_integral}}{=} -2 \int_{-\infty}^0 V^2 G_\infty( -V^2 ) dV =\\  
    \hspace{-0.6cm}= \frac{-\sqrt{\pi} e^{-\xi}}{4 \sqrt{2 \sigma^2 |\bar\eta| }} \begin{cases}
        2\sigma^2 I_{1/4}(\xi) + \bar\eta^2 \left[I_{1/4}(\xi) -I_{-1/4}(\xi)-I_{3/4}(\xi)+I_{5/4}(\xi)\right], \quad &\text{if } \bar\eta > 0,\\
        2\sigma^2 I_{1/4}(\xi) + \bar\eta^2 \left[ I_{1/4}(\xi) +I_{-1/4}(\xi)+I_{3/4}(\xi)+I_{5/4}(\xi)\right], &\text{if } \bar\eta < 0,
    \end{cases}
    \nonumber
\end{gather}
and $v_0=-\pi r_0$ if $\bar\eta = 0$. 

\begin{figure}[ht!]
    \centering{
    \includegraphics[width=7.5cm]{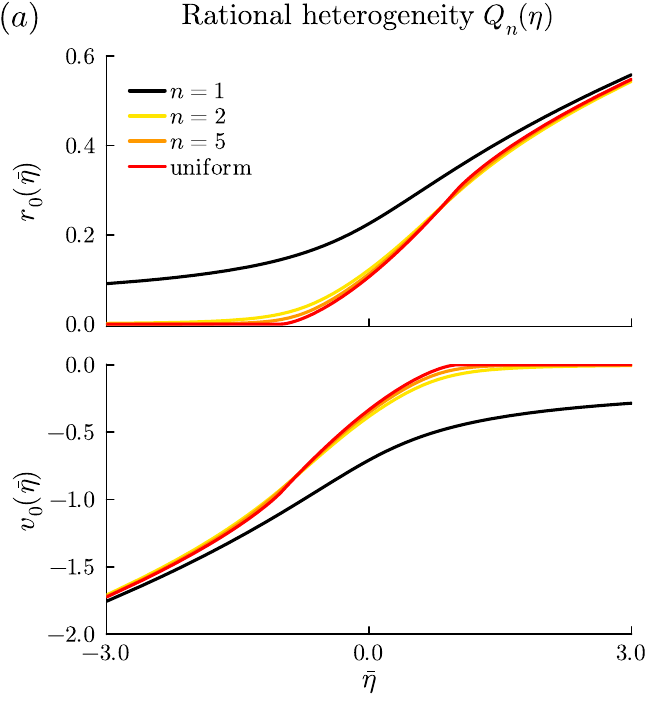}
    \includegraphics[width=7.5cm]{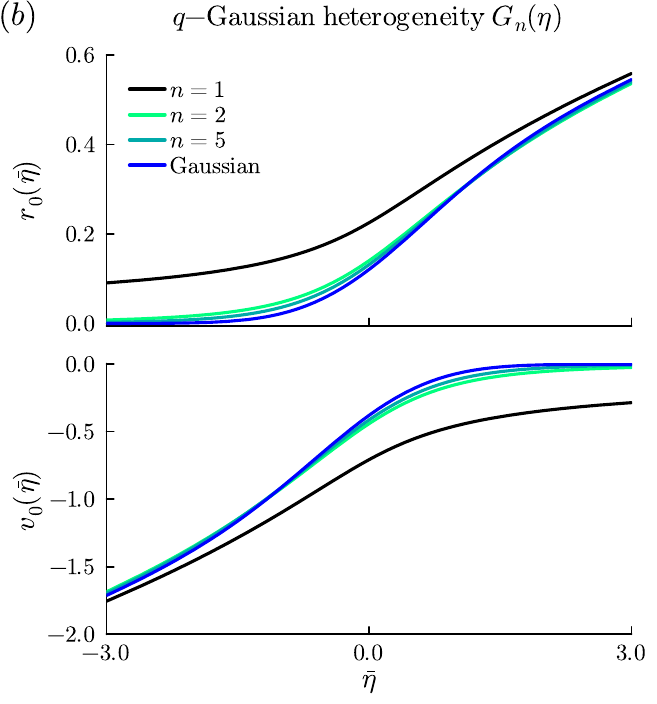}
    \caption{Stationary firing rates $r_0$ (top) and mean voltages $v_0$ (bottom) for heterogeneous populations of QIF neurons whose inputs follow (a) rational or (b) $q$-Gaussian distributions.
    For large excitatory (inhibitory) mean inputs $\bar\eta$, the firing rates (mean voltages) coincide with each other; otherwise,
    the fat tails of the Lorentzian distribution lead to devations from the values for rational or $q$-Gaussian heterogeneity, computed via \cref{eq:r0v0_Lorentzian,eq:rv0_Qn,eq:r0v0_Gn_bk}, as well as for uniform and Gaussian heterogeneities, computed via \cref{eq:rv0_uniform,eq:r0_Gauss,eq:v0_Gauss}.
    }
    }
    \label{fig:2}
\end{figure}

\subsection{The curious case of Lorentzian heterogeneity}
\label{subsec:RandV_Lorentzian}
While both the rational distributions $Q_n(\eta)$, \cref{eq:rv0_Qn}, as well as $q$-Gaussian distributions $G_n(\eta)$, \cref{eq:rv0_Gn}, lead to complex sums for the population firing rate and the mean voltage consisting of $n$ complex numbers,
the scenario dramatically simplifies in the case of a Lorentzian distribution $g_L(\eta) = \frac{1/\pi}{(\eta-\bar\eta)^2 + 1}$ with unit HWHM centered at $\bar\eta$, both \cref{eq:rv0_Qn,eq:rv0_Gn} yield with $n=1$ and the single pole $\eta = \bar\eta - i$ in the lower half-plane, 
\begin{equation}
    \pi r_0 - iv_0 = \int_{\mathbb{R}} \sqrt{\eta}g_L(\eta)d\eta =  \sqrt{\bar{\eta} - i}.
    \label{eq:rv0_Lorentzian}
\end{equation}
Computing the squared absolute value as well as the square of both sides of \cref{eq:rv0_Lorentzian}, leads to the relations
\begin{equation}\label{eq:r0v0_Lorentzian_relations}
    (\pi r_0)^2 + v_0^2 = \sqrt{\bar \eta^2 +1}, \quad (\pi r_0)^2 - v_0^2 = \bar \eta \quad \text{and} \quad 2 r_0 v_0 = -1/\pi \ . 
\end{equation}
We will come back to the last two relations later when dealing with the dynamic mean-field theory.
Now, from the second relation, we find the useful link $\pi r_0 = \sqrt{\bar\eta + v_0^2}$, where we used the positive root for physically meaningful, i.e.~non-negative, firing rates.
However, for $\bar\eta=0$, we have $v_0 = -\pi r_0$ which can be derived as follows:
Taking the sum and the difference of the first two relations in \cref{eq:r0v0_Lorentzian_relations},
results in explicit formulae for the population firing rate and the mean voltage
\begin{subequations}\label{eq:r0v0_Lorentzian}
    \begin{align}
        r_0 &= \frac1{\sqrt2 \pi} \sqrt{\bar\eta + \sqrt{\bar\eta^2 + 1}},\\
        v_0 &= \frac{-1}{\sqrt2} \sqrt{-\bar\eta + \sqrt{\bar\eta^2 + 1}}.
    \end{align}
\end{subequations}
The formulae \cref{eq:r0v0_Lorentzian} equally hold in the case of global coupling, where the mean $\bar\eta$ depends on the stationary firing rate $r_0$ and mean voltage $v_0$ as in \cref{eq:Udyn}.
Note that we could have obtained \cref{eq:r0v0_Lorentzian} also directly from decomposing the square root of a complex number into its real and imaginary parts, that is,
\[
    \pi r_0 - iv_0 = \sqrt{\bar{\eta} - i} = \frac{1}{\sqrt{2}} \left( \sqrt{\bar\eta + \sqrt{\bar\eta^2 + 1}} + i \sqrt{- \bar\eta + \sqrt{\bar\eta^2 + 1}} \right).
\]

\subsubsection{Firing rate distribution}
For Lorentzian distributed $\eta$, we find a concise formula for the firing rate distribution $P_0(f)$ from \cref{eq:firing_rate_dist} using the helpful relations~\eqref{eq:r0v0_Lorentzian_relations},
\begin{align*}
    P_0(f) &= \frac{2\pi f}{[(\pi f)^2 -\bar \eta]^2 + 1} = \frac{2\pi f }{(\pi f)^4 -2 (\pi f)^2\bar \eta + \bar \eta^2 + 1} \\
    &= \frac{ - 4\pi^2 r_0v_0 f}{(\pi f)^4 -2 (\pi f)^2 [(\pi r_0)^2 - v_0^2] + [(\pi r_0)^2 + v_0^2]^2} \\
    &= \frac{ - 4\pi^2 r_0v_0 f}{[(\pi f)^2 - (\pi r_0)^2]^2 +2 (\pi f v_0)^2 + 2(\pi r_0 v_0)^2 + v_0^4}
\end{align*}
With recurrent chemical coupling only ($J\neq 0 = g$), the equation simplifies and allows for averaging over the frequencies $f$ to obtain also the equation for the population firing rate as
\begin{align*}
    P_0(f) &= \frac{2\pi f}{[(\pi f)^2 -\bar \eta - Jr_0]^2 + 1}, \quad \text{where}\quad  \pi r_0 = \frac1{\sqrt{2}} \sqrt{\bar \eta + Jr_0 + \sqrt{(\bar \eta + Jr_0)^2 + 1}}.
\end{align*}

\subsubsection{Stationary voltage density for Lorentzian distributed parameters}
\label{subsubsec:Lorentzian_P0V}
We can also compute the total voltage density $P_0(V)$ explicitly by evaluating the integrals in \cref{eq:total_voltage_P0V},
\begin{align*}
    P_0(V) = \frac1\pi \int_0^\infty \frac{\sqrt{\eta} \ g_0(\eta)}{V^2 + \eta} d\eta + \int_{-\infty}^0 \delta(V+\sqrt{-\eta})g_0(\eta)d\eta.
\end{align*}
The first integral can be solved by, first, taking the Laplace transform and considering the limit $s\to0$ and, then, applying \cref{eq:r0v0_Lorentzian} together with the relations~\eqref{eq:r0v0_Lorentzian_relations}
\begin{align*}
    \frac{1}{\pi}\int_0^\infty \frac{\sqrt{\eta} \ g(\eta)}{V^2 + \eta} d\eta &= \frac{1}{\pi}\lim_{s\to 0} \int_0^\infty e^{-s\eta} \frac{\sqrt{\eta} }{V^2 + \eta} \frac{1/\pi}{1 + (\eta - \bar\eta)^2} d\eta\\
    &\hspace{-2.5cm}= \frac{i}{2\pi} \frac{\sqrt{-\bar\eta - i} -\sqrt{ -\bar\eta +i} }{ (\sqrt{-\bar\eta - i} +|V|) (\sqrt{-\bar\eta + i} + |V|)}
    = \frac{i}{2\pi} \frac{ (-v_0 - i\pi r_0) - (-v_0 + i\pi r_0) }{ |V|^2 - 2 |V|v_0^2 + \sqrt{\bar\eta^2 + 1}}\\
    &= \frac{r_0}{( |V| - v_0 )^2 + (\pi r_0)^2}
\end{align*}
The second integral as given in \cref{eq:total_voltage_2nd_integral} can be simplified with the relations \eqref{eq:r0v0_Lorentzian_relations} as
\begin{align}
    \int_{-\infty}^0 \delta(V+\sqrt{-\eta})g(\eta)d\eta &= \frac{2}\pi \frac{|V|}{ 1 + (\bar\eta + V^2)^2 } \times \Theta(-V)\nonumber\\
    &= \frac{4 r_0 v_0 |V| }{(2\pi r_0 v_0)^2 + [ V^2 - v_0^2 + (\pi r_0)^2]^2} \times \Theta(-V) \ . 
\end{align}
When considering the sum of both integrals for $V\ge0$ and $V<0$, it turns out that they coincide and yield an expression that is independent of the mean $\bar\eta$ and of the HWHM but solely determined by the firing rate and the mean voltage,
\begin{equation}\label{eq:P0V_Lorentzian}
    P_0(V) = \frac{r_0}{(V-v_0)^2 + \pi^2r_0^2},
\end{equation}
which is a Lorentzian distribution of $V$ with mean $v_0$ and half-width at half-maximum $\pi r_0 >0$.
Notably, the Lorentzian voltage distribution \eqref{eq:P0V_Lorentzian} of $V$ for Lorentzian heterogeneity of $\eta$ is independent of recurrent coupling and holds true for any $J\in\mathbb R$ and $g>0$.
In short, Lorentzian input distributions lead to Lorentzian voltage distributions in macroscopic stationary regimes. This is most astonishing and not at all the case for other input distributions, see \cref{fig:v_dist}, where obviously a Gaussian (uniform) input distribution does not lead to a Gaussian (uniform) voltage distribution.

The singularity of Lorentzian distributions in the context of heterogeneous QIF neurons manifests even more remarkably in the fact that, for Lorentzian input distributions,
the Lorentzian distribution of voltages $P(V,t)$ determines an attractive invariant manifold beyond macroscopic stationary regimes, as has been proven in~\cite{montbrio2015macroscopic,pietras2023exact}.
That is, given an initial distribution $P(V,0)$ of voltages $V$ with mean voltage $v(0)=v_0$ and firing rate $r(0)=r_0$, the distribution of voltages $V$ is given at all times $t>0$ by the Lorentzian
\begin{equation}\label{eq:PVt_Lorentzian}
    P(V,t) = \frac{r(t)}{[V-v(t)]^2 + \pi^2r(t)^2} \ 
\end{equation}
whose mean coincides with the mean voltage $v(t)$ and its width is given by $\pi r(t)$,
where $r(t)$ and $v(t)$ can now vary in time.
The long term behavior of QIF populations with Lorentzian distributed inputs is thus completely determined by the dynamics of the firing rate and the mean voltage, and vice versa. 
In \cite{montbrio2015macroscopic}, it was already shown that the dynamics of $r(t)$ and $v(t)$ can be explicitly written as two coupled ordinary differential equations.
In \cref{sec:DMF}, we will present these exact firing rate equations for heterogeneous populations of QIF neurons whose inputs follow rational or $q$-Gaussian distributions and, in \cref{sec:nonuniversal}, we will compare their respective collective dynamics with respect to generic and nonuniversal behavior.

\section{Dynamic mean-field theory}\label{sec:DMF}

\begin{figure}[ht!]
    \centering{
        \includegraphics[width=15cm]{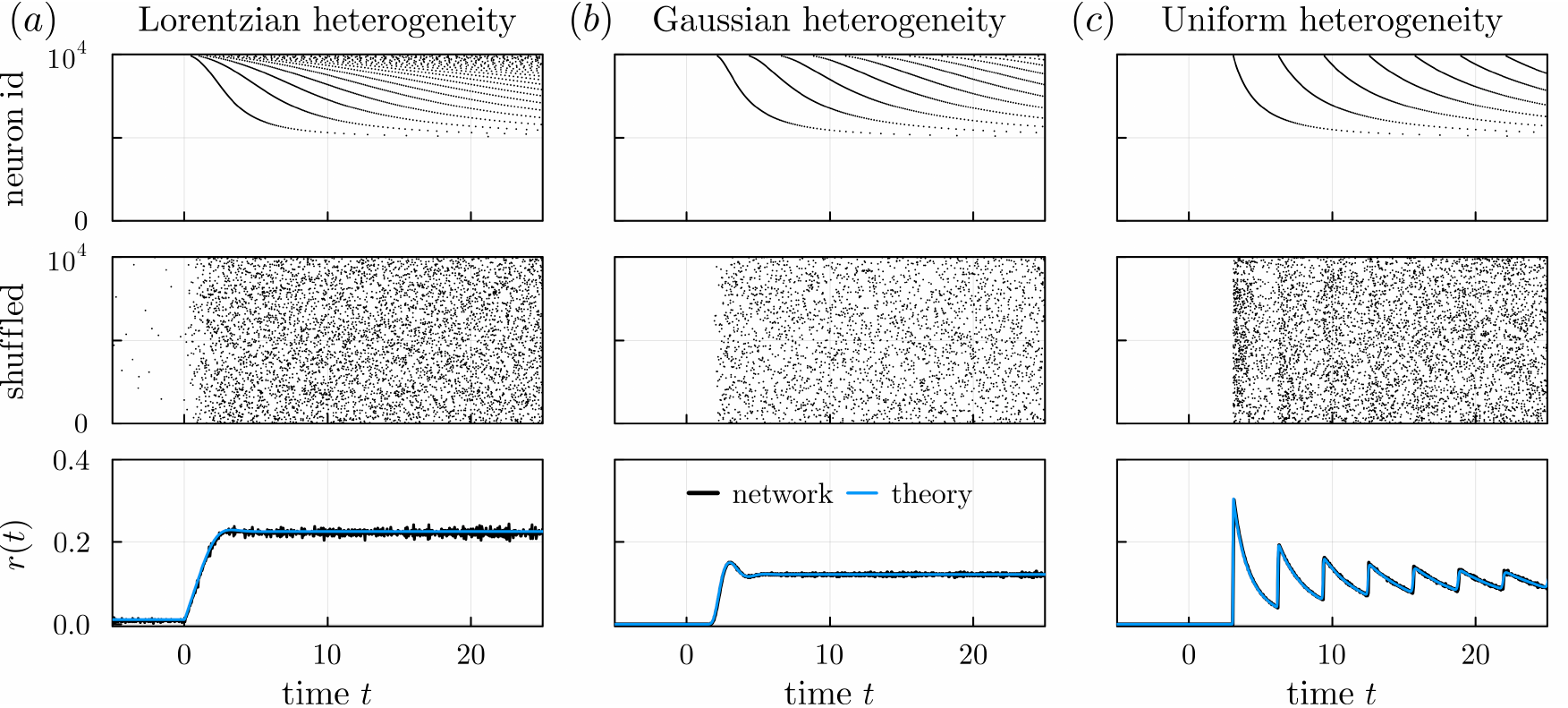}
        \caption{Spike synchrony of uncoupled heterogeneous populations of QIF neurons depends on the type of heterogeneity.
        We simulated \cref{eq:QIF} with $N=10.000$ neurons for (a) Lorentzian, (b) Gaussian and (c) uniform distributions of inputs $\eta_j$.
        Top: raster plots of the neurons' spike times $t_j^k$, ordered with respect to the value of $\eta_j$, with $\eta_1 < \dots < \eta_N$.
        Middle: raster plots with shuffled neuron ids, which makes synchronous firing better visible.
        Bottom: Population firing rate $r(t)$ computed for the network simulations via \eqref{firing_rate_formula} with $\tau_r=0.05$ and compared with the theory \cref{FRE_general,eq:FRE_rational,eq:W_multi} with $n=100$.
        Spike synchrony manifests as damped oscillations (``ringing''), see panel (c).
        The mean of the distributions $g_0(\eta)$ is $\bar\eta=-100$ for $t<0$ and $\bar\eta=0$ afterwards, the HWHM is $\Delta=1$, and simulation time step is $dt=5\times10^{-4}$.}
    }
    \label{fig:netsim_uncoupled}
\end{figure}

The analytic treatment put forward in \cref{sec:SMF} showed how the collective activity of globally coupled QIF neurons~\eqref{eq:QIF} with arbitrary heterogeneity of inputs $\eta_j$ can be determined in macroscopic stationary regimes, assuming that the population firing rate $r(t)=r_0$ and the mean voltage $v(t)=v_0$ are constant in time.
The approach above, however, does not allow one to determine the stability of these macroscopic fixed points, nor to track transient behavior.
This becomes important when the type of heterogeneity determines the degree of neuronal synchrony, which manifests in a ringing effect of damped oscillatory transients when neurons spike together (``spike synchrony''), see \cref{fig:netsim_uncoupled}.
For Lorentzian heterogeneity, the ringing effect is minimal, whereas for uniform heterogeneity damped oscillations are more pronounced. 

Towards a fully dynamic picture of the collective behavior, which not only accounts for the correct stability properties but also for the exact collective dynamics (see, e.g., the thin blue lines in \cref{fig:netsim}b), we allow that the mean input $\bar\eta=\bar\eta(t)$ depends on the dynamics of the mean fields as in \cref{eq:U_all}.
First, we consider QIF neurons~\eqref{eq:QIF} with excitability parameters $\eta$ that are drawn from a distribution density $g_t(\eta)$ with time-dependent mean $\bar\eta(t)$ and unit HWHM $\Delta=1$.
After presenting the dynamical mean-field theory for this simplified scenario, we will then obtain the collective dynamics in more realistic scenarios by reversing the nondimensionalization and rescaling as in \cref{nondimensionalization,rescaling} for the general QIF dynamics~\eqref{eq:QIF_all2all}.

The starting point for the dynamic mean-field theory is the so-called Lorentzian ansatz, which is related to the Ott-Antonsen ansatz~\cite{ott2008low} through a conformal mapping~\cite{montbrio2015macroscopic}.
In brief, we generalize the asymptotic conditional density~\eqref{eq:conditional_voltage_distribution} by imposing that, conditioned on the excitability parameter $\eta$, the voltages $V(t;\eta)$ of all those neurons that have excitability parameter $\eta$ are distributed according to a Lorentzian
\begin{equation}
    P(V,t;\eta) = \frac{1}{\pi} \frac{x(t;\eta)}{ [V-y(t;\eta)]^2 + x(t;\eta)^2}
    \label{Lorentzian_ansatz}
\end{equation}
with time-varying mean $y(t;\eta)$ and half-width $x(t;\eta)$ yet to be defined.
$P(V,t;\eta)$ conveniently includes the two asymptotic states of the QIF neuron~\cref{eq:QIF} described in \cref{subsec:firing,subsec:voltage}: the periodically firing one for $\eta>0$ with $y\to 0$ and $x\to f(\eta)$, as well as the excitable one for $\eta<0$, where $P(V,t;\eta)$ converges to a Dirac $\delta$-function with $x\to 0$ and $y\to -\sqrt{-\eta}$.
Moreover, the two time-dependent variables $x(t;\eta)$ and $y(t;\eta)$
have a clear biophysical meaning~\cite{montbrio2015macroscopic}: 
The firing rate $r(t;\eta)$ of neurons with a given $\eta$ is $r(t;\eta) = x(t;\eta)/\pi$.
The quanitity $y(\eta,t)$ corresponds to the mean of the membrane potential $y(t;\eta) = \mathrm{p.v.} \int_\mathbb{R} V\rho(V,t;\eta)dV$, where p.v.\ denotes the Cauchy principal value of the integral.
The population firing rate and the mean voltage are obtained by averaging over $\eta$,
\begin{equation}
    r(t) = \int_{\mathbb{R}} r(t;\eta)g_t(\eta) d\eta = \frac{1}{\pi} \int_{\mathbb{R}} x(t;\eta)g_t(\eta) d\eta \quad \text{and}\quad v(t) = \int_{\mathbb{R}} y(t;\eta)g_t(\eta) d\eta \ .
    \label{rv_to_wofeta}
\end{equation}
Relation~\eqref{eq10} motivates the introduction of the complex-valued variables $W(t)= \pi r(t) + iv(t)$ and $w(t;\eta)= \pi r(t;\eta) + iv(t;\eta)$, which are linked through 
\begin{equation}
    W(t)= \pi r(t) + iv(t) = \int_{-\infty}^\infty \big[x(t;\eta) + iy(t;\eta) \big]g_t(\eta) d\eta =  \int_{-\infty}^\infty w(t;\eta) g_t(\eta) d\eta.
\end{equation}
On the assumption that $w(t;\eta)$ is an analytic function in the complex-valued parameter $\eta$ in the lower complex half-plane and converges exponentially to zero as $\mathrm{Im}(\eta)\to-\infty$,
we can employ the Cauchy Residue Theorem to find
\begin{equation}
    W(t) = \pi r(t) + iv(t) = -2\pi i \ {\textstyle\sum_k} \mathrm{Res} (w(t;\eta)g_t(\eta); \zeta_k)
    \label{eq:W(t)}
\end{equation}
with the sum taken over the poles $\eta=\zeta_{k=1,2,\dots}$ of $g_t(\eta)$ in the lower half-plane---the same strategy has already been pursued in \cref{sec:SMF} and, although at this point it seems counterintuitive to introduce $W(t)$ as the complex conjugate of $\pi r_0 - iv_0$, \cref{eq10}, the dynamics of $W(t)$ will exhibit the same fixed point solutions as derived in \cref{sec:SMF}. 
The reason for choosing the poles in the lower half-plane boils down to the dynamics of $w(t;\eta)$ being restricted to the right complex-half plane, where $x(t; \eta) = \pi r(t; \eta) \ge 0$:
Indeed, the dynamics of $w(t;\eta)$ are obtained from the continuity equation that assures conservation of the number of QIF neurons \cref{eq:QIF},
\begin{equation}
    \partial_t P(V, t ;\eta) + \partial_V \big[ (V^2 + \eta) P \big] = 0,
\end{equation}
by plugging in the Lorentzian ansatz~\eqref{Lorentzian_ansatz};
the dynamics of $w(t; \eta)$ read~\cite{montbrio2015macroscopic,ratas2016macroscopic,pietras2019exact}
\begin{equation}
    \partial_t w(t;\eta) = \partial_t x(t;\eta) + i \partial_t y(t;\eta)= 2xy  + i( y^2 - x^2 + \eta) = i[\eta - w(t;\eta)^2 ] \ .
\end{equation}
As the Lorentzian ansatz \cref{Lorentzian_ansatz} describes a probability density, it cannot become negative and hence $x(t; \eta) \geq 0$ for all $t$ and $\eta$,
which is equivalent to requiring that biophysically meaningful firing rates $r(t;\eta) = x(t;\eta)/\pi \ge 0$ are non-negative.
For complex-valued $\eta=\eta_r + i\eta_i \in \mathbb{C}$, we thus require that $\partial_t \mathrm{Re}[w(t;\eta)]$ evaluated at $x(t;\eta)=0$ be positive, i.e.\ $\partial_t \mathrm{Re}[w(t;\eta)] \big|_{x=0} = - \eta_i>0$. 
Consequently, the imaginary part of $\eta$ must be negative, $\mathrm{Im}(\eta) < 0$,
and the integral in \cref{eq:W(t)} has to be evaluated by closing the contour in the lower half-plane.

The dynamics of $W(t)$, and thus of $r(t)$ and $v(t)$, then follow immediately.
If all the $\zeta_{k=1,2,\dots}$ are simple poles of $g_t(\eta)$ in the lower half plane, such as in the case of rational functions $Q_n(\eta)$, then
\begin{subequations}
    \label{eq:W_simple}
    \begin{align}
    W(t) &= \pi r(t) + iv(t) = -2\pi i\ {\textstyle\sum_k} \mathrm{Res} (g(\eta); \zeta_k) W_k(t),\quad \text{where}\\
    \dot W_k &= i [\zeta_k - W_k^2], \quad k=1,2,\dots
    \end{align}
\end{subequations}
Note that the poles $\zeta_k = \zeta_k(t)$ depend on the time-dependent mean $\bar\eta(t)$ and incorporate also the information about the unit HWHM of the distribution $g_t(\eta)$.
If the poles are of higher order, as is the case, e.g., for $q$-Gaussians $Q_n(\eta)$, the dynamics of $W(t)$ has to be found from
\begin{subequations}
    \label{eq:W_multiple}
    \begin{align}
    W(t) &=  -2\pi i \sum_k \frac{1}{(n_k-1)!} \lim_{\eta \to \zeta_k} \frac{d^{n_k-1}}{d\eta^{n_k-1}} \Big( (\eta-\zeta_k)^{n_k} g_t(\eta)w(t;\eta) \Big)\ ,
    \end{align}
where the $k$-th pole $\zeta_k$ is of order $n_k \ge 1$.
Often, $W(t)$ can still be expressed as a linear combination of time-dependent order parameters in the form
\begin{align}
    W(t) = \sum_k b_k W_k(t), \quad \text{where} \quad W_k(t) = \frac{\beta_k}{(k-1)!} \lim_{\eta\to\zeta_k} \frac{\partial^{k-1}}{d\eta^{k-1}} w(t;\eta)
\end{align}
for some parameters $b_k, \beta_k$.
The typical mean fields---firing rate $r(t)$ and mean voltage $v(t)$---are then recovered from $W(t)$ as 
\begin{align}
    r(t) = \tfrac{1}\pi \mathrm{Re}[W(t)] \quad \text{and} \quad v(t) = \mathrm{Im}[W(t)] \ .
\end{align}
\end{subequations}
Needless to say, the collective dynamics put forward in this section naturally reproduce the stationary firing rate and mean voltage solutions from \cref{sec:SMF}. 
What is more, the dynamical mean-field theory goes beyond stationary regimes and also captures transient dynamics and synchronous regimes, as will be the focus below.

\subsection{Collective dynamics of heterogeneous populations}
Before investigating concrete examples, we first summarize the exact low-dimensional collective dynamics for Lorentzian, rational and $q$-Gaussian distributions $g_t(\eta)$. 
For Lorentzian heterogeneity, the $W_k$-dynamics dramatically simplifies to two-dimensional ``firing rate equations'' that explicitly govern the dynamics of the firing rate $r(t)$ and mean voltage $v(t)$. 
For other types of heterogeneity, the mean-field reduction, and thus the collective dynamics, is higher-dimensional.

\subsubsection{Lorentzian distribution and firing rate equations}
For a Lorentzian distribution of inputs $\eta$, $g_t(\eta)$ has a single simple pole in the lower half-plane, $\eta=\zeta_-  = \bar\eta + \exp[-i\pi/2]=\bar\eta - i$, and \cref{eq:W_simple} with \cref{eq:lemma_residue_rational} dramatically simplifies as
\begin{align}
    W(t) = \pi r(t) + iv(t) = -2\pi i \frac{-1}{2\pi} \sin(\pi/2) (-i) W_1(t) = -i^2 W_1(t) = W_1(t).
\end{align}
The dynamics of $W=W_1$ reads $\dot W= 1 + i [ \bar\eta(t) - W(t)^2]$,
which can conveniently be expressed in terms of $r(t)$ and $v(t)$ as the {\bf firing rate equations} (FRE)~\cite{montbrio2015macroscopic,pietras2019exact,guerreiro2022exact}
\begin{subequations}\label{FRE_general}
    \begin{align}
        \dot r &= 1/\pi + 2 rv,\\
        \dot v &= v^2 - (\pi r)^2 + \bar\eta.
    \end{align}
\end{subequations}

\subsubsection{Rational distributions}
For rational distributions, $g_t(\eta)=Q_n(\eta)$, the collective dynamics in terms of the complex variable $W(t)$ follows analogously to \cref{eq:rv0_Qn} as 
\begin{subequations}\label{eq:FRE_rational}
    \begin{align}
    &W(t) = \tfrac{\pi}{a} r(t) +iv(t) = i \sin\left(\frac{\pi}{2n}\right) \sum_{k=1}^n \alpha_k W_k(t),\quad \text{where} \quad \alpha_k = e^{-i\pi(2k-1)/(2n)}, \label{eq:FRE_rational1}\\
    &\dot W_k = i [\bar\eta + \alpha_k - W_k^2 ] \quad \text{for} \quad k=1,2,\dots,n.
    \end{align}
\end{subequations}
The stationary solutions of \cref{eq:FRE_rational} satisfy $W_k \equiv \sqrt{\bar\eta + \alpha_k} = \sqrt{\zeta_k}$, which together with \cref{eq:FRE_rational1} yield stationary solutions identical to those given by \cref{eq:zetak_rational,eq:rv0_Qn}.
The dynamic mean-field dynamics \cref{eq:FRE_rational} is thus consistent with the stationary mean-field theory of \cref{sec:SMF}.

\subsubsection{$q$-Gaussian distributions}
For $q$-Gaussian distributions, $g_t(\eta)=G_n(\eta)$, the collective dynamics in terms of the complex variable $W(t)$ can be obtained as in~\cite{pyragas2022mean,pyragas2023effect} similar to our derivation of~\cref{eq:r0v0_Gn}, so that 
\begin{subequations}\label{eq:W_multi}
    \begin{align}
    &W(t) = \pi r(t) +iv(t) = \sum_{k=1}^n b_k W_k(t),\quad \text{where}\\
    &\dot W_1 = i \big[\bar\eta-W_1^2\big] + \beta_n^{-1/2}, \\
    &\dot W_2 = - 2i W_1W_2  - \beta_n^{-1/2} , \\
    &\dot W_k = - i\sum_{l=1}^k  W_{k-l+1}W_l ,\quad k=3,\dots,n,
    \end{align}
\end{subequations}
with $\beta_n = 2^{1/n}-1$ and the $b_k$ are given through the recurrence relation \cref{eq:r0v0_Gn_bk}, that is, $b_1 = 1, b_k = b_{k-1} (n+1-k)/(n-k/2) $ for $k=2,\dots,n$.
We leave it as a straightforward exercise to the interested reader to show that the stationary solutions of \cref{eq:W_multi} for $q$-Gaussian distributions are consistent with the stationary mean-field theory of \cref{sec:SMF}.

\section{Nonuniversal collective dynamics of heterogeneous inhibitory QIF neurons
}
\label{sec:nonuniversal}
For the concrete example of the QIF dynamics \cref{eq:QIF_all2all} with global coupling via gap junctions and with first-order chemical synapses, 
\begin{align*}
    \tau_m \dot V_j &=a V_j^2 + a\Delta \xi_j + I(t) + J \tau_m s(t) + g [v(t) - V_j], \\
    \tau_s \dot s &= - s + r(t),
\end{align*}
the collective dynamics for Lorentzian, rational and $q$-Gaussian heterogeneity $g(\xi)$ are readily obtained by reversing the shift and rescaling of variables and the normalization of \cref{transformation}.
Recall that the individual inputs $\xi_j$ follow normalized distribution densities with zero mean and unit HFHM, so that the parameter $\Delta>0$ scales the overall heterogeneity.
The mean field $W(t)$ introduced in \cref{sec:DMF} corresponds here to the mean field of the $\tilde U$-variables in \cref{eq:U_all},
that is, one should read $\tilde{\mathcal Y}(\tilde t) = \pi \tilde r + i \tilde u$ with $\tilde u = \langle \tilde U_j\rangle = \tilde v - \tilde g/2$.
Reshifting $\tilde{\mathcal Y}$ and rescaling yields 
\begin{subequations}
    \begin{align}
    &\tilde W = \tilde{\mathcal Y} + i\tilde g/2 \quad \text{as well as}\quad \tilde W_k = \tilde{\mathcal Y}_k + i\tilde g/2, \quad \text{and}\\
    &W = \sqrt{\Delta} \tilde{W} \quad \text{and}\quad W_k = \sqrt{\Delta} \tilde{W}_k \quad \text{for}\quad k=1,2,\dots,n.
    \end{align}
\end{subequations}
By applying the other backtransformations of \cref{transformation} 
to the dynamics~\cref{FRE_general,eq:FRE_rational,eq:W_multi} 
and setting $a=1$ without loss of generality 
(alternatively, $a$ can be included in the time constant $\tau_m$), 
we obtain the {\bf Lorentzian firing rate equations (FRE)}
\begin{subequations}\label{eq:FRE_all}
    \begin{align}
        \tau_m \dot r&= \nicefrac{\Delta}{(\pi\tau_m)} + 2 rv - gr,\\
        \tau_m \dot v &= 
        v^2 - (\pi \tau_m r)^2 + I(t) + J\tau_m s(t),\\
        \tau_s \dot s &= -s + r,
    \end{align}   
\end{subequations}
the {\bf rational collective dynamics} (with $n>2$)
\begin{subequations}\label{eq:W_simple_all}
    \begin{align}
    &W(t) = \pi \tau_m r(t) + iv(t) = i \sin\left(\frac{\pi}{2n}\right) \sum_{k=1}^n \alpha_k W_k(t),\quad \text{where} \quad \alpha_k=e^{-i\pi \frac{2k-1}{2n}},\\
    &\tau_m \dot W(t) = i \big[ I(t) + \Delta\alpha_k + J \tau_m s(t) -W_k^2\big] + g [iv(t) - W_k] , \quad k=1,2,\dots, n ,\\
    &\tau_s \dot s= -s + r,
    \end{align}
\end{subequations}
and the {\bf $\boldsymbol{q}$-Gaussian collective dynamics} (with $n>2$)
\begin{subequations}\label{eq:W_multi_all}
    \begin{align}
    W(t) &= \pi \tau_m r(t) +iv(t) = \sum_{k=1}^n b_k W_k(t),\quad \text{where}\\
    \tau_m \dot W_1 &= i [ I(t) + J\tau_m s(t) - W_1^2]  + g [ i v(t) -  W_1] + \Delta\beta_n^{-1/2}, \\
    \tau_m \dot W_2 &= -g W_2 -i 2W_1W_2  - \Delta\beta_n^{-1/2} , \\
    \tau_m \dot W_k &= -gW_k -i\sum_{l=1}^k W_{k-l+1}W_l ,\quad k=3,\dots,n,\\
    \tau_s \dot s&= -s + r,
    \end{align}
\end{subequations}
with $\beta_n = 2^{1/n}-1$ and $b_1 = 1, b_k = b_{k-1} (n+1-k)/(n-k/2) $ for $k=2,\dots,n$ as before.

In the following, we will investigate in which way the collective dynamics \cref{eq:FRE_all,eq:W_simple_all,eq:W_multi_all} coincide in the case of only inhibitory coupling with first-order chemical synapses (\cref{subsec:chem_only}), of only electrical-coupling via gap junctions (\cref{subsec:gap_only}), and of both electrical and chemical coupling (\cref{subsec:gap_and_chem}).
As we demonstrated that the network dynamics are exactly captured by the dynamic mean-field theory, see \cref{fig:netsim,fig:netsim_uncoupled}, we will concentrate on bifurcation analysis of the collective dynamics and compare the resulting phase diagrams.

\subsection{Chemical coupling only $(J<0,g=0)$}
\label{subsec:chem_only}
\begin{figure}[!t]
    \centering{
    \includegraphics[width=7.5cm]{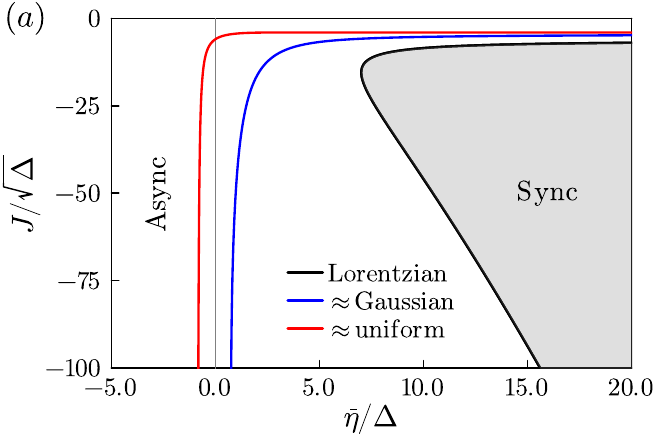}
    \includegraphics[width=7.5cm]{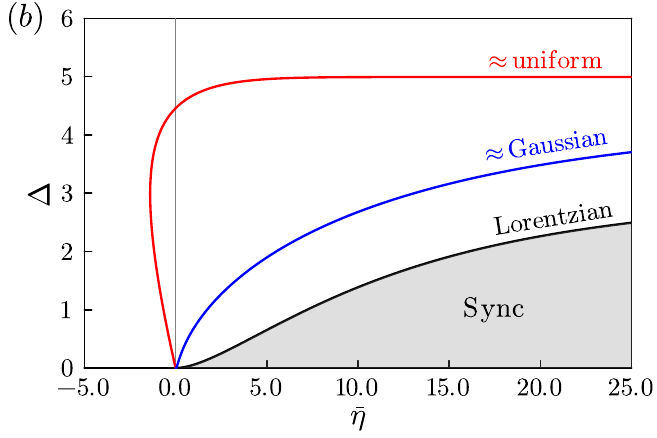}
    \caption{Synchronization transitions from an asynchronous state (Async) to collective oscillations (Sync) occur via supercritical Hopf bifurcations in networks of inhibitory QIF neurons with fast first-order synaptic kinetics and Lorentzian (black), Gaussian (blue) or uniform (red) heterogeneity.
    Phase diagrams for (a) the mean input $\bar\eta/\Delta$ versus strength of inhibition $J/\sqrt\Delta$ rescaled with respect to the degree of heterogeneity $\Delta$, and for (b) $\bar\eta$ versus $\Delta$ at a fixed level of inhibition, $J=-20$. 
    For Lorentzian and Gaussian heterogeneity, collective oscillations are only possible for excitatory mean inputs $\bar\eta>0$, whereas QIF neurons with uniform heterogeneity exhibit a diversity-induced transition to synchrony for slightly negative $\bar\eta$ upon increasing $\Delta$, see panel (b). 
    Bifurcation boundaries were obtained with AUTO using rational and $q$-Gaussian distribution functions $Q_n,G_n$ with $n=20$ for uniform and Gaussian heterogeneity and $n=1$ for Lorentzian heterogeneity.
    }
    \label{fig:INH}
    }
\end{figure}

How heterogeneity influences a population of QIF neurons with global and instantaneous chemical coupling, was already analyzed in the appendix of \cite{montbrio2015macroscopic}.
That collective oscillations are not possible in inhibitory networks with instantaneous chemical synapses, but require the presence of synaptic kinetics, was proven for Lorentzian heterogeneity in \cite{devalle2017firing,coombes2018next,ratas2016macroscopic}; a similar picture seems to persist for rational and $q$-Gaussian heterogeneity, see \cite{pyragas2022mean,pyragas2023effect}.
In \cref{fig:INH}, we compare the transitions to collective oscillations, or synchrony, for different types of heterogeneity and with fast first-order synaptic kinetics (the synaptic time constant is $\tau_s = 5$ms$=\tau_m/2$).
In particular, we consider the exact collective dynamics \cref{eq:FRE_all} for Lorentzian heterogeneity and approximate the collective dynamics for Gaussian and uniform heterogeneity by the $q$-Gaussian and rational collective dynamics \cref{eq:W_simple_all,eq:W_multi_all}, respectively, with $n=20$, which presents an accurate fit according to \cref{fig:1}.

For Lorentzian heterogeneity (black curves), there is a triangular-like region of synchrony (the shaded ``Sync'' region in \cref{fig:INH}a) that requires strong input $\bar\eta \gg 1$ and sufficient inhibition $J \ll -1$. 
For Gaussian (blue) and uniform (red) heterogeneity, the Sync-region is stretched to the left; the mean input necessary to induce collective oscillations drastically shrinks, $\bar\eta \approx 0$, and becomes almost independent of the strength of inhibition $J$. 
Curiously, for uniform heterogeneity, strong enough inhibition allows for collective oscillations even for slightly negative mean inputs.
In all three cases, the transitions to synchrony are  supercritical Hopf bifurcations, which we then analyzed upon varying the degree of heterogeneity $\Delta$. 
Although the regions of synchrony are rather different for Lorentzian and Gaussian heterogeneity, for a fixed value of inhibition ($J=-20$ in \cref{fig:INH}b) both types react quite similarly upon increasing $\Delta$.
The uniform distribution, by contrast, exhibits nonstandard behavior:
for slightly negative mean inputs $-1 < \bar\eta <0$ and upon increasing the width parameter $\Delta$ from zero, there is a transition from an asynchronous regime to collective oscillations (at $\Delta \approx 1$) and then again to asynchrony (at $\Delta \approx 4$). 
It seems, however, that this heterogeneity- or ``diversity-induced transition to synchrony'' is owed to the somewhat special properties of uniform heterogeneity, cf., e.g., its firing rate and voltage distributions in~\cref{fig:fr_dist,fig:v_dist}.

\subsection{Electrical coupling only $(g>0,J=0)$}
\label{subsec:gap_only}
\begin{figure}[!t]
    \centering{
    \includegraphics[width=7.5cm]{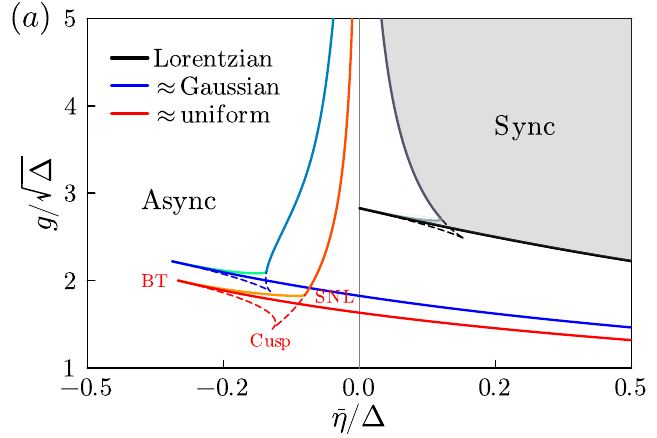}
    \includegraphics[width=7.5cm]{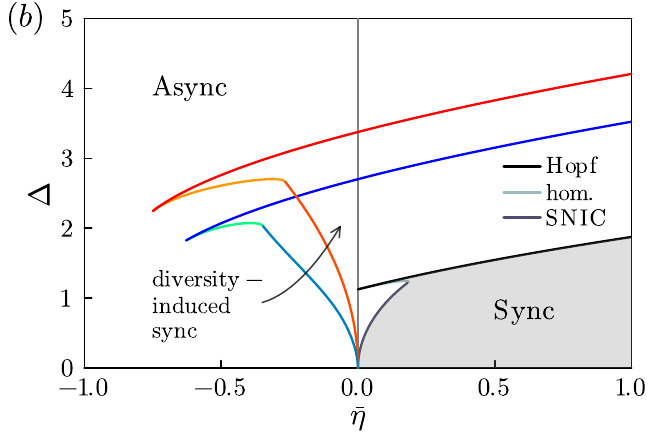}
    \caption{Synchronization transitions in networks of QIF neurons with electrical synapses (gap junctions) and Lorentzian (black), Gaussian (blue) or uniform (red) heterogeneity. 
    (a) In the phase diagram for rescaled mean input $\bar\eta/\Delta$ versus rescaled gap junction strength $g/\sqrt\Delta$, the Sync-region of collective oscillations (in the upper-right part) is bounded from below by a supercritical Hopf bifurcation (dark) that terminates for decreasing $\bar\eta$ in the codimension-2 Bogdanov-Takens bifurcation point (BT). 
    From there, a homoclinic (light) and a saddle-node bifurcation (dashed) emerge, which meet in a saddle-node-separatrix-loop (SNL) bifurcation point after the saddle-node bifurcation line has undergone a Cusp bifurcation. Above the SNL point, the saddle-node bifurcation occurs on the limit cycle, also known as a SNIC bifurcation, which asymptotically converges to $\bar\eta=0$ for increasing gap junction strength $g\gg0$. 
    For Lorentzian heterogeneity, the Sync-region lies completely in the right-half plane and collective oscillations require positive mean inputs $\bar\eta>0$. 
    For Gaussian and uniform heterogeneity, the SNIC boundary moves to the left of the $g$-axis and collective oscillations become possible for negative mean inputs $\bar\eta<0$.
    (b) For fixed gap junction-strength $g=3$, the phase diagram for mean input $\bar\eta$ versus degree of heterogeneity $\Delta$ reveals a diversity-induced transition inside the Sync-region with negative mean input $\bar\eta<0$ and upon increasing $\Delta$ for Gaussian and uniform, but not for Lorentzian heterogeneity.
    Bifurcation boundaries were obtained with AUTO using rational and $q$-Gaussian distribution functions $Q_n,G_n$ with $n=20$ for uniform and Gaussian heterogeneity and $n=1$ for Lorentzian heterogeneity.}
    \label{fig:GJ}
    }
\end{figure}

The phenomenon of {\bf diversity-induced synchronization} becomes generic in heterogeneous networks of QIF neurons when they are coupled via electrical synapses in form of gap junctions (\cref{fig:GJ}).
Collective dynamics for gap junction-coupled QIF neurons with Lorentzian heterogeneity have been analyzed in detail in \cite{pietras2019exact}, where a synchronization region (bounded by a Hopf bifurcation from below, a SNIC bifurcation from the left, and a homoclinic bifurcation between them) was reported for sufficiently strong gap junctions $g>0$ and excitatory mean inputs $\bar\eta>0$.
Thus, at least half the neurons must be self-oscillatory to set up a collective rhythm across the network, independent of the electrical coupling strength $g$.
For Gaussian and uniform heterogeneity, by contrast, collective oscillations are also possible if only a minority of the QIF neurons are self-oscillatory: synchrony becomes possible for negative mean inputs $\bar\eta$ as the SNIC boundaries in \cref{fig:GJ}(a) move to the left of the $g$-axis.

The diversity-induced transition to synchrony becomes visible when fixing a certain level of gap junction strength ($g=3$ in \cref{fig:GJ}b) and increasing the degree of heterogeneity $\Delta$.
While the SNIC boundary for Lorentzian heterogeneity moves strictly right from the origin, those for Gaussian and uniform heterogeneity move to the left. 
That is, for negative mean inputs, say $\bar\eta =-0.2$ and small heterogeneity, the networks with Gaussian and uniform heterogeneity are first in an asynchronous regime. 
By increasing $\Delta$, and thus making the network more diverse, we move up above the SNIC boundary and inside the synchronization region, where collective oscillations emerge with arbitrary small frequency and finite amplitude. Increasing $\Delta$ even more across the Hopf bifurcation boundary (almost perpendicular to the SNIC one), makes the collective oscillations cease with finite frequency and arbitrary small amplitude.
For Lorentzian heterogeneity, such a diversity-induced transition is not possible: synchrony either exists for arbitrary small $\Delta$ and then ceases for larger $\Delta$ (if $\bar\eta>0$), or synchrony does not exist at all (if $\bar\eta<0$).
For gap junction-coupled networks of QIF neurons, hence, Lorentzian heterogeneity exhibits nonuniversal collective behavior.

\subsection{Electrical and chemical coupling $(g>0,J<0)$}
\label{subsec:gap_and_chem}

\begin{figure}[!t]
    \centering{
    \includegraphics[width=7.5cm]{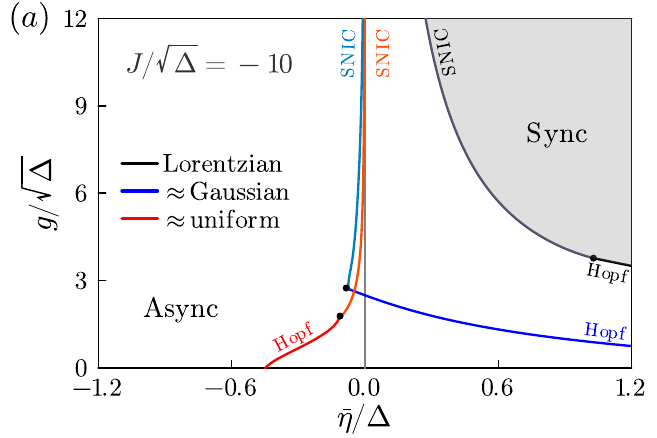}
    \includegraphics[width=7.5cm]{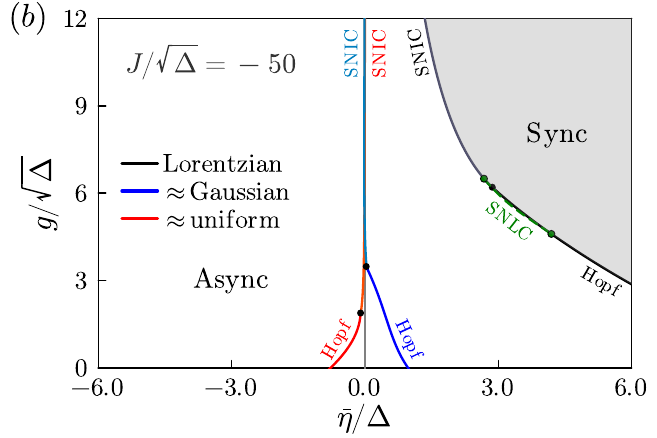}
    \caption{Synchronization transitions in networks of inhibitory QIF neurons with fast first-order synaptic kinetics ($\tau_s=\tau_m/2=5$ms), electrical synapses (gap junctions) and Lorentzian (black), Gaussian (blue) or uniform (red) heterogeneity. 
    For a fixed level of inhibition, the Sync-region of collective oscillations lies in the upper-right part of the $(\bar\eta/\Delta)$-versus-$(g/\sqrt\Delta)$-phase diagram similar to \cref{fig:GJ}(a). 
    QIF neurons with Lorentzian heterogeneity always require positive mean inputs $\bar\eta>0$ to synchronize. 
    (a) For moderate inhibition, $J/\sqrt\Delta=-10$, 
    the SNIC bifurcation boundaries, which delimit the Sync-region on the left, move to the left-half plane ($\bar\eta<0$) for Gaussian and uniform, but not for Lorentzian heterogeneity.
    While the supercritical Hopf bifurcations emanate from the black dots towards larger mean inputs $\bar\eta$ for Gaussian and Lorentzian heterogeneity, for uniform heterogeneity the Hopf curve runs to the left and further enlarges the Sync-region of collective oscillations.
    (b) For strong inhibition, $J/\sqrt\Delta=-50$, 
    the SNIC bifurcation boundaries for Gaussian and uniform heterogeneity almost coincide with the $g$-axis.
    While the supercritical Hopf bifurcation for Gaussian heterogeneity emanates towards the bottom right, the one for uniform heterogeneity continues to run to the left.
    Hence, only in case of uniform, but neither for Gaussian nor Lorentzian, heterogeneity can the combination of strong inhibition and gap junction coupling enable collective oscillations in QIF networks where the majority of the neurons are excitable and not self-oscillatory. 
    Besides, for Lorentzian heterogeneity, there is a small parameter range that allows for a subcritical Hopf bifurcation---and, thus, enables bistability between an asynchronous and a synchronous state---close to the right of the saddle-node of limit cycles (SNLC) bifurcation (between the two green dots).
    All the black dots cover the drastically shrunk triangular bifurcation structure of BT, Cusp and SNL points of \cref{fig:GJ}(a).
    Bifurcation boundaries were obtained with AUTO using rational and $q$-Gaussian distribution functions $Q_n,G_n$ with $n=20$ for uniform and Gaussian heterogeneity as before.}
    \label{fig:GJINH}
    }
\end{figure}

In networks of inhibitory QIF neurons without gap junctions, Lorentzian heterogeneity yielded apparently generic collective dynamics.
By contrast, in gap junction-coupled networks of QIF neurons without chemical synapses, Lorentzian heterogeneity resulted in rather nonuniversal collective dynamics. 
So, what kind of behavior will the Lorentzian firing rate equations produce in the presence of both inhibitory synapses and gap junctions?
To answer this, we investigated the collective dynamics in \cref{fig:GJINH} for (a) moderate and (b) high levels of inhibition.
In all cases, the Sync-region of collective oscillations is bounded mainly by a Hopf and a SNIC bifurcation. 
For Lorentzian heterogeneity, collective oscillations always require significant excitatory mean input $\bar\eta >0$; the stronger inhibition, the stronger must the excitatory drive be. 
For moderate inhibition, Gaussian and uniform heterogeneity allow for collective oscillations with slightly negative mean inputs $\bar\eta<0$; for Gaussian heterogeneity, however, the gap junction strength $g$ needs to be sufficiently high.
Notably, for weak gap junction coupling, collective oscillations are only possible with uniform heterogeneity. 

The bifurcation scenario changes for strong inhibition (\cref{fig:GJINH}b), where now also Gaussian heterogeneity allows for collective oscillations with weak gap junction coupling. 
However, this comes at the cost that collective oscillations for Gaussian heterogeneity now always require excitatory drive $\bar\eta>0$, as is the case for Lorentzian heterogeneity.
Intriguingly, the Hopf bifurcation boundary for Lorentzian heterogeneity exhibits a change of criticality:
for weak inhibition, the Hopf bifurcation is always supercritical, whereas for strong inhibition, the Hopf bifurcation becomes subcritical (at the right green dot in \cref{fig:GJINH}b). 
This change of criticality gives rise to bistability between an asynchronous state and collective oscillations, so that a slight increase of the mean input now allows for a swifter and more rapid synchronization compared to the conventional transitions to synchrony.
Still, this phenomenon is exclusive to Lorentzian heterogeneity and limited to only a small parameter region (close to the green-dashed SNLC line in \cref{fig:GJINH}b), hence it is questionable whether this behavior of QIF neurons with Lorentzian heterogeneity is indeed generic.

\section{Discussion \& conclusion}
\label{sec:conclusion}

The derivation of exact, low-dimensional models for heterogeneous populations of QIF neurons is generally made assuming Lorentzian heterogeneity.
While other types of heterogeneity admit exact mean-field reductions, the resulting mean-field models are higher-dimensional and more challenging to analyze \cite{klinshov2021reduction,pyragas2021dynamics,pyragas2022mean,pyragas2023effect}.
Here, we started from first principles and uncovered the mathematical relations why Lorentzian distributed inputs result in Lorentzian output (the voltage distribution), at least in macroscopic stationary regimes (\cref{sec:SMF}).
The two defining parameters---mean and width---of the Lorentzian voltage distribution directly correspond to the mean voltage and the population firing rate.
Such a striking correspondence between input/output distributions and biophysically relevant macroscopic quanitities is exclusive to Lorentzian heterogeneity
and
may hint at some deeper truths why the Lorentzian voltage distribution is a globally attracting and invariant manifold of the collective dynamics of globally coupled QIF neurons with Lorentzian distributed inputs \cite{pietras2016ott,engelbrecht2020ott,pietras2023exact}.

By formalizing and generalizing the ideas in Appendix C of \cite{montbrio2015macroscopic}, 
we first analyzed voltage and firing rate distributions in macroscopic stationary regimes of heterogeneous populations of QIF neurons with arbitrary input distributions, which yielded an alternative approach to compute steady states of the population firing rate, see also \cite{buice2013dynamic,laing2014derivation}.
We then focused on rational and $q$-Gaussian distributions that interpolate between Lorentzian and uniform or between Lorentzian and Gaussian distributions, respectively.
These distributions are in so far advantageous as they have finitely many poles in the complex plane and, thus, allow for a convenient computation of stationary population firing rates and mean voltages by means of Cauchy's residue theorem (\cref{subsec:RandV_Ldistributions}).
Lorentzians only have two poles, whereas for $q$-Gaussian and rational distributions the number of poles quickly increases, especially when approximating Gaussian or uniform distributions, 
which makes the computation of the mean fields significantly more involved.
This picture also carries over to the dynamic mean-field theory in \cref{sec:DMF} that goes beyond macroscopic stationary states.
For Lorentzian heterogeneity, two coupled ordinary differential equations---the so-called firing rate equations---govern the dynamics of the population firing rate and the mean voltage, which exactly capture the collective dynamics of the full network.
For $q$-Gaussian and rational heterogeneity, many more than two coupled differential equations are necessary to capture the collective dynamics.

In \cref{sec:nonuniversal} we studied the different mean-field models with particular focus on how the different types of heterogeneity affect the transitions to synchrony and collective oscillations.
In other words, we asked whether the collective dynamics were generic across different types of heterogeneity, or whether one particular type gave rise to nonuniversal behavior.
Bifurcation analyses led us to the conclusion that the Lorentzian firing rate equations are a good proxy for the collective dynamics of inhibitory QIF neurons with other types of heterogeneity and away from critical bifurcation boundaries.
While the parameter region for collective oscillations enlarges, e.g., for Gaussian compared to Lorentzian heterogeneity, see also \cite{pyragas2022mean,clusella2024exact}, the differences are rather quantitative but not qualitative.

However, when QIF neurons are coupled with gap junctions, Lorentzian heterogeneity leads to nonuniversal behavior as it does not allow for diversity-induced transitions to synchrony, which generically occur for other types of heterogeneity when the mean input is slightly negative and the degree of heterogeneity is constantly increased.
The corresponding phase diagram (\cref{fig:GJ}b) strongly resembles the scenario of globally coupled active rotators, see Fig.~4 in \cite{lafuerza2010nonuniversal}, where caution was advised against the use of Lorentzian distributions in ensembles of excitable systems. 
In the end, one always has to find a compromise between accuracy (of the more convoluted collective dynamics for rational or $q$-Gaussian heterogeneity) and convenience, as the Lorentzian firing rate equations are lowest-dimensional and more amenable to analysis but may not be the best description for the collective dynamics of QIF networks with uniform or Gaussian heterogeneity.

Eventually, next-generation neural mass models have been shown to capture the collective dynamics not only of heterogeneous, but also of noisy QIF neurons. 
To be more precise, the firing rate equations for heterogeneous networks of deterministic QIF neurons with Lorentzian distributed inputs are identical to those of a homogeneous network where QIF neurons are driven by independent Cauchy noise~\cite{clusella2024exact,pietras2024pulse,pietras2023exact,pyragas2023effect}.
This equivalence seems to be limited again to the curious case of the (Cauchy-)Lorentzian distribution and cannot be generalized, e.g., to Gaussian heterogeneity vis-\`a-vis Gaussian noise~\cite{clusella2024exact}.
A first step towards a comprehensive investigation of the collective behavior for stochastic QIF neurons with different types of noise has already been achieved \cite{goldobin2024chaos}, but crucial steps are still missing.
For instance, even in macroscopic stationary regimes, it is unclear whether the statistics of interspike intervals---the stochastic equivalent to firing rate distributions in heterogeneous networks of deterministic QIF neurons---can be obtained in closed form. 
It thus remains to be seen whether a general dynamic mean-field theory for networks of noisy QIF neurons can finally be obtained.

\section*{Acknowledgments}
B.P. has received funding from the European Union's Horizon 2020 research and innovation programme under the Marie Sk{\l}odowska-Curie Grant No. 101032806.
E.M. acknowledges support by the Agencia Estatal de Investigaci\'on under the Project No. PID2019-109918GB-I00, and by the Generalitat de Catalunya under the grant 2021 SGR0 1522 646.

\bibliographystyle{siamplain}

\end{document}